\documentclass[reprint,prl,footinbib,floatfix,twocolumn,longbibliography]{revtex4-2}
\usepackage{times}
\usepackage{amssymb}
\usepackage{color}
\usepackage{amsmath}
\usepackage{amsbsy}
\usepackage{amsthm}
\usepackage{graphicx}
\usepackage{bbm}
\usepackage{bm}
\usepackage{epsfig}
\usepackage{xfrac}
\usepackage{xcolor}
\usepackage{enumerate}
\usepackage{multirow}
\usepackage{physics}
\usepackage[T2A,T1]{fontenc}
\usepackage{ tipa }
\usepackage[english]{babel}
\usepackage{symbols}
\usepackage{makecell}
\usepackage{appendix}
\usepackage{dsfont}
\usepackage{float}
\usepackage{pdfpages}
\usepackage{bbold}
\usepackage{placeins}
\usepackage{relsize}
\usepackage{xcolor}
\usepackage{scalerel}
\usepackage{tikz}
\usetikzlibrary{svg.path}

\DeclareMathSymbol{\shortminus}{\mathbin}{AMSa}{"39}

\definecolor{orcidlogocol}{HTML}{A6CE39}
\tikzset{
  orcidlogo/.pic={
    \fill[orcidlogocol] svg{M256,128c0,70.7-57.3,128-128,128C57.3,256,0,198.7,0,128C0,57.3,57.3,0,128,0C198.7,0,256,57.3,256,128z};
    \fill[white] svg{M86.3,186.2H70.9V79.1h15.4v48.4V186.2z}
                 svg{M108.9,79.1h41.6c39.6,0,57,28.3,57,53.6c0,27.5-21.5,53.6-56.8,53.6h-41.8V79.1z M124.3,172.4h24.5c34.9,0,42.9-26.5,42.9-39.7c0-21.5-13.7-39.7-43.7-39.7h-23.7V172.4z}
                 svg{M88.7,56.8c0,5.5-4.5,10.1-10.1,10.1c-5.6,0-10.1-4.6-10.1-10.1c0-5.6,4.5-10.1,10.1-10.1C84.2,46.7,88.7,51.3,88.7,56.8z};
  }
}

\newcommand\orcid[1]{\href{https://orcid.org/#1}{$\,$\mbox{\scalerel*{
\begin{tikzpicture}[yscale=-1,transform shape]
\pic{orcidlogo};
\end{tikzpicture}
}{|}}}}

\definecolor{myurlcolor}{rgb}{0.0,0.39,0.0}
\definecolor{myrefcolor}{rgb}{0.0,0.39,0.0}

\usepackage[colorlinks]{hyperref}
\hypersetup{unicode=true,
    bookmarksopen=false,
    breaklinks=false,
    pdfborder={0 0 0},
	bookmarksnumbered=false,
	pdfstartview={FitH},
	citecolor={myurlcolor},
	linkcolor={myrefcolor},
	urlcolor={myurlcolor}}

\usepackage{soul}

\makeatletter
\AtBeginDocument{\let\LS@rot\@undefined}
\makeatother 

\makeatletter
\def\maketitle{
\@author@finish
\title@column\titleblock@produce
\suppressfloats[t]}
\makeatother

\begin{document}

\title{Optimal Phase-Insensitive Force Sensing with Non-Gaussian States}

\author{Piotr~T.~Grochowski\orcid{0000-0002-9654-4824}}
\email{piotr.grochowski@upol.cz}
\author{Radim~Filip\orcid{0000-0003-4114-6068}}
\email{filip@optics.upol.cz}
\affiliation{Department of Optics, \href{https://ror.org/04qxnmv42}{Palacký University}, 17. listopadu 1192/12, 771 46 Olomouc, Czech Republic}

\begin{abstract}
Quantum metrology enables sensitivity to approach the limits set by fundamental physical laws.
Even a single continuous mode offers enhanced precision, with the improvement scaling with its occupation number. 
Due to their high information capacity, continuous modes allow for the engineering of quantum non-Gaussian states, which not only improve metrological performance but can also be tailored to specific experimental platforms and conditions.
Recent advancements in control over continuous platforms operating in the quantum regime have renewed interest in sensing weak forces, also coupling to massive macroscopic objects.
In this work, we investigate a force-sensing scheme where a physical process completely randomizes the direction of the induced phase-space displacement, and the unknown force strength is inferred through excitation-number-resolving measurements.
We find that $N$-spaced states, where only every $N^{\text{th}}$ Fock state occupation is nonzero, approach the achievable sensing bound.
Additionally, non-Gaussian states are shown to be more resilient against decoherence than their Gaussian counterparts with the same occupation number.
While Fock states typically offer the best protection against decoherence, we uncover a transition in the metrological landscape---revealed through a tailored decoherence-aware Fisher-information-based reward functional---where experimental constraints favor a family of number-squeezed Schrödinger cat states.
Specifically, by implementing quantum optimal control in a minimal spin-boson system, we identify these states as maximizing force sensitivity under lossy dynamics and finite system controllability.
Our results provide a pathway for enhancing force sensing in a variety of continuous quantum systems, ranging from massive systems like mechanical oscillators to massless systems such as quantum light and microwave resonators.
\end{abstract}

\maketitle

\textit{Introduction.}---Increasing measurement precision is vital for progress in fundamental physics and emerging technologies.
It enables the discovery of new phenomena~\cite{Safronova2018,Chou2023}, tests of established theories~\cite{Tse2019,VirgoCollaboration2019,Karr2020}, and the development of ultra-sensitive sensors~\cite{Ye2008,deAngelis2008,Aslam2023}.
A central goal in quantum metrology is to surpass the standard quantum limit (SQL), where measurement noise reaches the vacuum level~\cite{Paris2009,Degen2017,Pirandola2018}.
This can be achieved with either entangled multi-probe states or nonclassical multi-excitation states of a single continuous mode~\cite{Duivenvoorden2017,Braun2018,Fadel2025}.
The latter has advanced rapidly across optical~\cite{Rouviere2024}, microwave superconducting~\cite{Pan2025}, atomic~\cite{Facon2016,Burd2019,Gilmore2021,Delakouras2023}, and nanomechanical platforms~\cite{Rakhubovsky2024,Satzinger2018,Mason2019,Wollack2022,vonLupke2022,Bild2023,Youssefi2023,Millen2020,Gonzalez-Ballestero2021}, each providing new opportunities to probe quantum limits and enhance precision sensing.

\begin{figure}[ht!]
    \includegraphics[width=\linewidth]{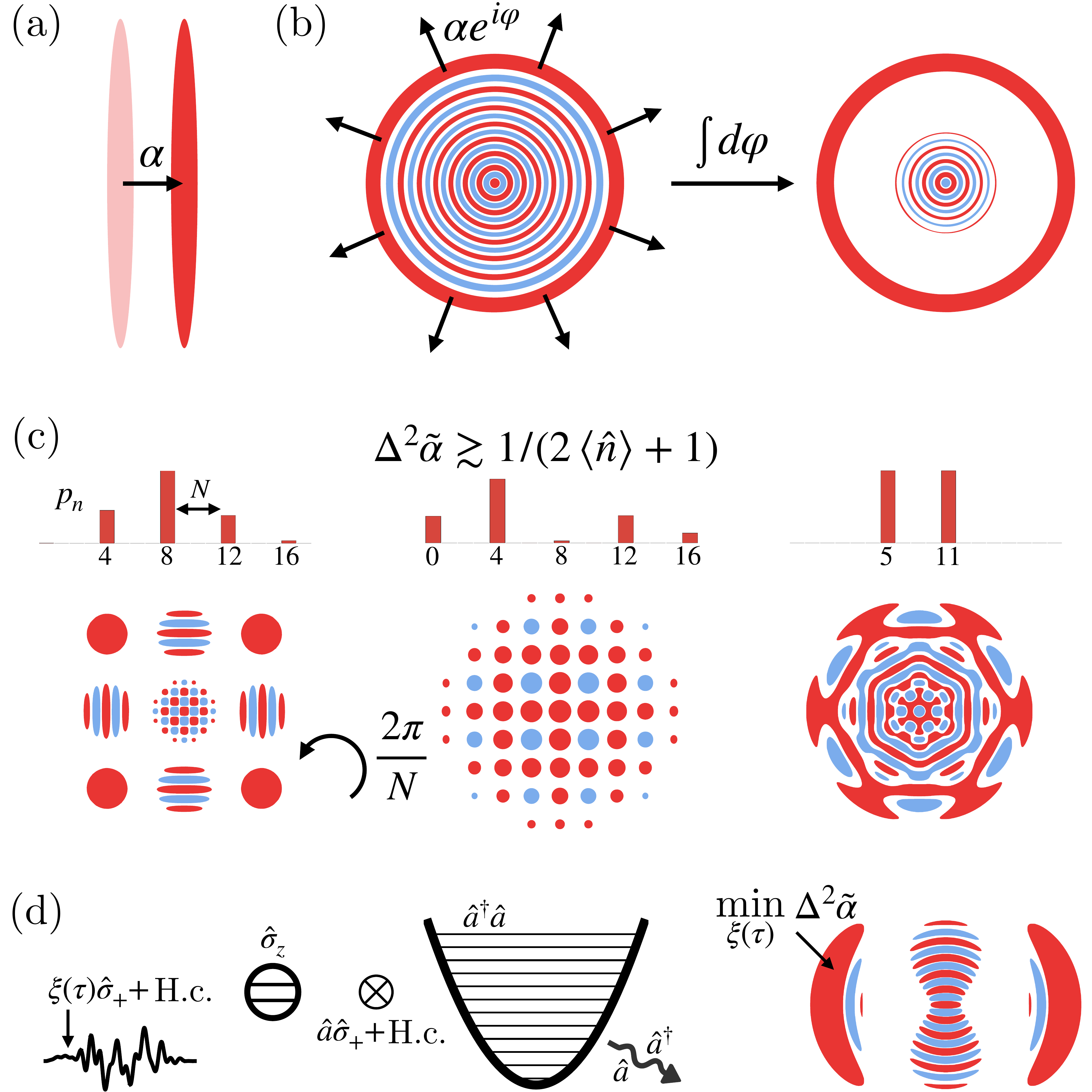}  
    \caption{(a) Squeezed, narrow phase-space features along the known force direction are necessary for sensing schemes to manifest quantum advantage. (b) If the direction of phase displacement is completely randomized during the sensing process, the narrow features need to be present along all directions. (c) Quantum non-Gaussian states with discrete rotational symmetry~\cite{Grimsmo2020}, such as compass~\cite{Shukla2023}, grid~\cite{Gottesman2001a}, or $\ket{mn}$ states~\cite{Asenbeck2025,Lachman2025} (from left to right), approach the Heisenberg limit with excitation-number-resolving measurements. (d) In an optimally controlled spin-boson model with finite controllability and decoherence, a family of number-squeezed cat states maximizes achievable sensitivity for weak force sensing.
  \label{Fig1}}
\end{figure}

For force sensing, precision beyond the SQL relies on squeezing phase-space features below the vacuum level.
Demonstrations include preparing bosonic systems in nonclassical states~\cite{Lachman2022,Walschaers2021,Rakhubovsky2024,Frigerio2025}, such as cat states~\cite{Vlastakis2013,Pan2025}, Fock states~\cite{Chu2018,Wolf2019,Podhora2022,Hofheinz2008,Eickbusch2022,Deng2024,Rahman2025}, and their superpositions~\cite{McCormick2019,Wang2019,Kovalenko2025}.
Typically, maximal advantage requires alignment of the prepared state with the force direction~\cite{Hempel2013,Lo2015}.
When alignment is unavailable, states with small sub-Planck features in all phase-space directions---such as Fock~\cite{Wolf2019,Deng2024} or grid states~\cite{Valahu2025}---offer a workaround.
Here we examine the scenario of completely randomized displacement phase~\cite{Gorecki2022}, relevant to trapped ions~\cite{Wolf2019,Deng2024,Valahu2025}, solid-state oscillators~\cite{Bild2023,Rahman2025}, gravitational-wave detection~\cite{Ballantini2003,Carney2025}, and dark matter searches~\cite{Teufel2009,Backes2021,Brady2022,Higgins2024}.
In this setting, the task is to estimate the displacement amplitude $\PhaseSpaceDispAmp$ with high precision using the standard sequence of probe preparation, evolution under the sensing Hamiltonian for a time $\tau$, and measurement of the excitation number, while minimizing the number of repetitions.

We expand the known families of probe states that saturate the Heisenberg limit~\cite{Braunstein1994,Giovannetti2004,Maccone2020} to include those with discrete rotational phase-space symmetry~\cite{Grimsmo2020}.
Comparing different classes, we find that their sensitivity degrades differently under decoherence, with Fock states typically offering the best noise protection.
To model realistic preparation scenarios with finite controllability and intrinsic decoherence, we implement an optimal-control protocol in a minimal spin–boson model controlled solely via the spin~\cite{Liu2024}.
This yields a transition in the metrological landscape, where different non-Gaussian families emerge as optimal under varying decoherence levels and force amplitudes.
Our results extend to other nonlinear models, enabling metrological gain under realistic constraints.

\textit{Metrological gain of non-Gaussian states.}---We start with a single continuous-variable bosonic mode, where $\Creation$ adds a single excitation, and the canonical commutation relation $[\Annihilation,\Creation]=1$ holds.
We focus on sensing phase-space displacement, $\Displacement{\PhaseSpaceDisp} = \Displacement{\PhaseSpaceDispAmp e^{\ImagUnit \DispPhase}} = \exp (\PhaseSpaceDisp \Creation - \PhaseSpaceDisp^{\ast} \Annihilation)$, caused by an external classical force.
We are interested in processes where the displacement phase $\DispPhase$ is completely randomized, implying uniform classical mixing,
\begin{align}
    \DensityMatrix_\DispLetter^{i} =\int_{0}^{2 \pi} \frac{\dd \DispPhase}{2 \pi} \ \Unitary_{i}\rounds{\PhaseSpaceDispAmp,\DispPhase}
    \DensityMatrix \ \Unitary_{i}^{\dagger}\rounds{\PhaseSpaceDispAmp,\DispPhase}.
    \label{mixing}
\end{align}
We consider various physical realizations of such a channel,
\begin{align}
    \Unitary_{1} &= \Displacement{\PhaseSpaceDispAmp e^{\ImagUnit \DispPhase}}, \  \Unitary_{2} = \Displacement{\PhaseSpaceDispAmp} \Rotation(\DispPhase)
    \label{channels}
\end{align}
where $\Rotation(\DispPhase) = \exp(- \ImagUnit \DispPhase \Creation \Annihilation)$.
Here, $\Unitary_{1}$ is relevant for, e.g., trapped-ion systems~\cite{Oh2020} and  is a type of \textit{noisy spreading} channel~\cite{Gorecki2022}, while $\Unitary_{2}$ can describe phase-space rotation of the sensing state.
In such processes, the phase varies from experimental shot to shot, a case not equivalent to a scenario where the phase is fixed but unknown during the experiment~\cite{Fadel2025}.
After calculation, the diagonal elements of $\DensityMatrix_\DispLetter^{i}$, $\Prob_n^{\DispLetter} = \Tr( \ket{n}\bra{n} \DensityMatrix_{\DispLetter}^{i}  )$ are equal for $i=1,2$ and found to be~\cite{deOliveira1990}
\begin{align}
\Prob_n^{\DispLetter} = \sum_{k=0}^{\infty} \DensityMatrixElement_{k} e^{-\PhaseSpaceDispAmp^2} \rounds{\frac{n!}{k!}}^{\frac{k-n}{|k-n|}} \PhaseSpaceDispAmp^{2 |k-n|} \squares{ L_{\mu}^{|k-n|} \rounds{\PhaseSpaceDispAmp^2}}^2, \label{probneq}
\end{align}
where $  \DensityMatrix = \sum_{k,l} \DensityMatrixElement_{kl} \ket{k}\bra{l}$, $\mu = \text{min} (n,k)$, $L_{p}^{q}$ are associated Laguerre polynomials, and we have defined $\DensityMatrixElement_{k} \equiv \DensityMatrixElement_{kk}$ for simplicity.
On the other hand, the off-diagonal elements differ between the phase-randomization realizations.
For the protocol, we consider excitation number-resolving measurements, realized in systems with well-controlled spin-motion coupling~\cite{Meekhof1996,Chu2018,Podhora2022}, optical setups~\cite{Hadfield2009}, and possibly accessible only via nonharmonic potentials~\cite{Weiss2019,Grochowski2025a}.
Note that the additional phase mixing, induced by, e.g., phase noise, after the action of the channel~\eqref{mixing}, fully erases the phase and does not affect $\Prob_n^{\DispLetter}$, making such a scheme optimal.
To assess sensitivity, we use the Cram\'er-Rao bound~\cite{Holevo2011}, $\Delta \tilde{\PhaseSpaceDispAmp} \geq 1/ (\sqrt{M} \sqrt{\Fisher_{\DispLetter}})$,
where $\Delta \tilde{\PhaseSpaceDispAmp}$ is the root-mean-square error for the $\PhaseSpaceDispAmp$-estimator $\tilde{\PhaseSpaceDispAmp}$, $M$ is the number of experimental repetitions, and the Fisher information (FI) for projection on Fock states $\ket{n}\bra{n}$ is
\begin{align}
    \Fisher_{\DispLetter} = \sum_{n=0}^{\infty} \frac{1}{\Prob_n^{\DispLetter}} \rounds{\frac{\partial \Prob_n^{\DispLetter}}{\partial {\DispLetter} }}^2.
\end{align}
Note that $\Fisher_{\DispLetter}$ does not depend on the purity $\Tr \DensityMatrix^2$ of the probe state $\DensityMatrix$, as its off-diagonal terms do not contribute to the FI.
Because the number of repetitions $M$ varies across realizations and is often limited by technical limitations or available classical resources~\cite{Demkowicz-Dobrzanski2015,Hradil2019}, we focus on maximizing the asymptotic single-shot FI.

The classical limit for a sensing task is usually defined via a quantum FI maximized over all classical (with positive Glauber P-representation) states, which has been shown to equal 4 and is obtained by a pure coherent state~\cite{Wolf2019}.
In our case, the excitation-resolving measurement is optimal only for the ground state, yielding $\Fisher_{\DispLetter,0} = 4$~\cite{Oh2020,Ritboon2022}.
It allows us to use it as a reference value for the \textit{metrological gain} $\fisher_{\PhaseSpaceDispAmp}$, which we define as $\fisher_{\PhaseSpaceDispAmp} = \Fisher_{\PhaseSpaceDispAmp} / \Fisher_{\DispLetter,0}$~\cite{Wolf2019}.
It can be shown that the FI is upper-bounded by the occupation number~\cite{Wolf2019,Gorecki2022,Fadel2025},
\begin{align}
    \fisher_{\PhaseSpaceDispAmp} \leq  1 + 2 \angles{\NumberOperator},
    \label{bound}
\end{align}
where $\angles{\NumberOperator^k} = \Tr(\DensityMatrix \NumberOperator^k)$, $\NumberOperator = \Creation \Annihilation$, and Fock states $\ket{n}$ saturate this bound, $\fisher_{\PhaseSpaceDispAmp} (\ket{n}) = 1 + 2 n$~\cite{Wolf2019,Oh2020,Gorecki2022,Ritboon2022}.
Note that we do not aim to find an optimal measurement and compute the quantum FI for every considered probe state, but rather to compare the strategy involving excitation-resolving measurement with the upper bound~\eqref{bound}.

Let us introduce \textit{$N$-spaced} states, i.e., states where only every $N^{\text{th}}$ Fock state occupation is nonzero, 
\begin{align}
    \DensityMatrixElement_n \neq 0 \ \ \text{only if} \ \ \text{mod}(n,N)=n_0,
\end{align}
where $n_0 < N$.
By Taylor-expanding $\fisher_{\PhaseSpaceDispAmp}$ around $\PhaseSpaceDispAmp = 0$, we find that they approach the bound~\eqref{bound} for small ${\PhaseSpaceDispAmp}$ (see derivation in App.~\hyperref[appA]{A}): 
\begin{align}
    \fisher_{\PhaseSpaceDispAmp} &= 1 + 2 \angles{\NumberOperator}  - \mathcal{O}(\PhaseSpaceDispAmp^{2\lfloor N/2 \rfloor} )
    \label{N-scaling}    
\end{align}
for $N \geq 2$.
States that are 1-spaced do not exhibit metrological gain, with $\fisher_{\PhaseSpaceDispAmp} = \mathcal{O}(\PhaseSpaceDispAmp^2)$.
For 2-spaced states, i.e., states with well-defined parity, the subleading term is (cf. App.~\hyperref[appB]{B}):
\begin{align}
    \fisher_{\PhaseSpaceDispAmp} = 1 + 2 \angles{\NumberOperator} - 2 \PhaseSpaceDispAmp^2 \rounds{1 +  \angles{\NumberOperator} + \angles{\NumberOperator^2}} + \mathcal{O}(\PhaseSpaceDispAmp^4 ).
    \label{parity-scaling}
\end{align}
Pure $N$-spaced states exhibit discrete rotational symmetry in phase space, as eigenstates of $\exp(\ImagUnit 2\pi \NumberOperator / N)$.
This symmetry allows us to identify several families of metrologically useful states, beyond Fock states and squeezed vacuum~\cite{Gorecki2022}.
These families include multi-legged cat states~\cite{Zurek2001}, grid states~\cite{Gottesman2001a}, number-phase states~\cite{Grimsmo2020}, $\ket{mn}$ states~\cite{Kovalenko2025}, generalized squeezed states~\cite{Bazavan2024}, and others.
The metrological gain arises from $N$-spacedness itself, while the phase-space symmetry of pure $N$-spaced states is lost under the phase randomization considered here.
The gain coming from stronger number squeezing scales as $\mathcal{O}(\PhaseSpaceDispAmp^2)$, becoming significant only at larger $\PhaseSpaceDispAmp$ [cf. Fig.~\ref{Fig2} and Eq.~\eqref{parity-scaling}].
The differing scaling of 1- and 2-spaced states suggests that metrological gain is vulnerable to imperfections and decoherence~\cite{Kolodynski2010,Knysh2011}.
We compare $\fisher_\PhaseSpaceDispAmp$ for these different families in App.~\hyperref[appC]{C} (more details can be found in Supplemental Materials (SM)~\cite{SuppMat4}).

\textit{Gain under sensing-stage errors.}---Imperfections and decoherence can occur at any stage of sensing: preparation, probing, or measurement. 
Here we consider the first two, while the last can be improved through smart strategies~\cite{Ritboon2022} or hardware refinement.
To model noisy dynamics during the sensing stage, we use a Gaussian decoherence channel given by the Lindblad master equation for a mode coupled to a bath with thermal occupation $\ThermalOccupation$ and linear coupling $\PhotonLossCoeff$~\cite{Breuer2007},
\begin{align}
    \partial_{\Time} \DensityMatrix (\Time) &=  \sum_{i} \frac{1}{2} \squares{2 \Coll_{i} \DensityMatrix(\Time) \Coll_{i}^{\dagger} - \DensityMatrix(\Time ) \Coll_{i}^{\dagger} \Coll_{i} - \Coll_{i}^{\dagger} \Coll_{i} \DensityMatrix(\Time )   }, \nonumber \\
    \Coll_{1} &= \sqrt{\PhotonLossCoeff \rounds{1+\ThermalOccupation}} \Annihilation, \ \ \Coll_{2} =  \sqrt{\PhotonLossCoeff \ThermalOccupation} \Creation,
    \label{full-master}
\end{align}
where we work in a frame rotating with oscillator's free dynamics, $\Time$ is dimensionless time, and collapse operators $\Coll_{1}$ and $\Coll_{2}$ correspond to deexcitation and excitation, respectively, and $\DensityMatrix (\Time = 0)$ has been prepared for the purpose of sensing.
Such a model describes decoherence in mechanical systems, both clamped~\cite{Yang2024} and levitated~\cite{Gonzalez-Ballestero2019}, and in optical or microwave cavities~\cite{Liu2024}.
We approximate the joint action of noisy evolution and displacement by applying the displacement after the noisy dynamics, since it only weakly perturbs the state and may represent either a continuous force or a momentum kick.
We aim to study the decoherence susceptibility of the probe state, so we consider the limit $\SmallTimeBar = \PhotonLossCoeff \ThermalOccupation \Time \ll 1$~\cite{Fujii2013}, yielding
\begin{align}
\DensityMatrix (\Time) \approx \DensityMatrix\rounds{1   - \SmallTime \EffThermalMinus}   -  \SmallTime \EffThermalBar  \curlies{\NumberOperator,\DensityMatrix} + \SmallTime  \EffThermalMinus \Creation \DensityMatrix \Annihilation + \SmallTime  \EffThermalPlus \Annihilation \DensityMatrix \Creation,
\label{small-time}
\end{align}
where $\SmallTime = \PhotonLossCoeff \Time$, $\EffThermalBar = \ThermalOccupation + 1/2$, and $\EffThermal_{\pm} = \EffThermalBar \pm 1/2 $, and $\{A,B\}=AB+AB$.
We focus on two opposing regimes---pure loss ($\ThermalOccupation = 0$, symbol $\LossLetter$) and heating ($\ThermalOccupation \gg 1$, $\HeatLetter$).
The expansions of the FI around $\PhaseSpaceDispAmp = 0$ and $\SmallTime (\SmallTimeBar)= 0$ read (cf. App.~\hyperref[appD]{D})
\begin{align}
\fisher_\PhaseSpaceDispAmp^{\LossLetter} &= \angles{\NumberOperator} +1 + \frac{\angles{\NumberOperator}}{1+\PhotonLossCoeff \Time / \PhaseSpaceDispAmp^2} +  \mathcal{O}(\PhotonLossCoeff \Time \PhaseSpaceDispAmp^2), \label{perturbative}\\
\fisher_\PhaseSpaceDispAmp^{\HeatLetter} &= \frac{1+2\angles{\NumberOperator}}{1+\PhotonLossCoeff \ThermalOccupation\Time/\PhaseSpaceDispAmp^2} + \mathcal{O}(\PhotonLossCoeff \ThermalOccupation\Time \PhaseSpaceDispAmp^2),
\label{fixed_eps}
\end{align}
which holds for at least 3- [Eq.~\eqref{perturbative}] and 2-spaced [Eq.~\eqref{fixed_eps}] states.
Expressions~\eqref{perturbative} and~\eqref{fixed_eps} allow us to identify regimes, $\PhotonLossCoeff \Time /\PhaseSpaceDispAmp^2 \ll 1$ and $\PhotonLossCoeff \Time \ThermalOccupation / \PhaseSpaceDispAmp^2 \ll 1$, where the saturation of the bound~\eqref{bound} can be achieved in presence of decoherence~\cite{Kolodynski2010,Knysh2011}.
2-spaced states in the presence of loss exhibit similar behavior, however, not following exactly Eq.~\eqref{perturbative} (see App.~\hyperref[appD]{D}).
Analysis beyond the scaling of Eqs.~\eqref{perturbative} and~\eqref{fixed_eps}, both using Eq.~\eqref{small-time} [Figs.~\ref{Fig2}(a,b)] and full numerical solution of Eq.~\eqref{full-master} [Figs.~\ref{Fig2}(c,d)], reveal that different families of states exhibit different decoherence susceptibilities.
Fig.~\ref{Fig2} presents a comparison between Fock and Gaussian states (squeezed vacuum states)---identified as the most and least robust to decoherence among the considered families, respectively---where the latter are characterized by a number distribution
\begin{align}
    \DensityMatrixElement_n^\text{G} =
      \binom{n}{n/2} \frac{\angles{\NumberOperator}^{n/2}}{2^n \rounds{1+\angles{\NumberOperator}}^{(n+1)/2}} \ \  \text{if $n$ is even, else 0}. \label{gaussian}
\end{align}
Moreover, the dynamical range~\cite{Pan2025}, defined as the set of $\PhaseSpaceDispAmp$ values for which $\fisher_{\PhaseSpaceDispAmp} > 1$, is finite for Gaussian states and decreases with increasing decoherence.
In contrast, for Fock states, it persists across all tested $\PhaseSpaceDispAmp$.
In the metrologically most relevant regime of sub-vacuum displacements ($\PhaseSpaceDispAmp < 0.5$), both Fock and Gaussian states achieve FI above the SQL [cf. Fig.~\ref{Fig2}(e,f)].
However, while the FI of Gaussian states exhibits a sharp peak followed by rapid decay, the FI of Fock states remains nearly constant throughout this interval.
This advantage is not universal: in the heating case and at larger $\PhaseSpaceDispAmp$, some states may provide superior decoherence protection (see Supplemental Material~\cite{SuppMat4}).
Fock states were also found optimal under decoherence for stochastic force estimation~\cite{Gardner2025}.

\begin{figure}[t!]
    \includegraphics[width=\linewidth]{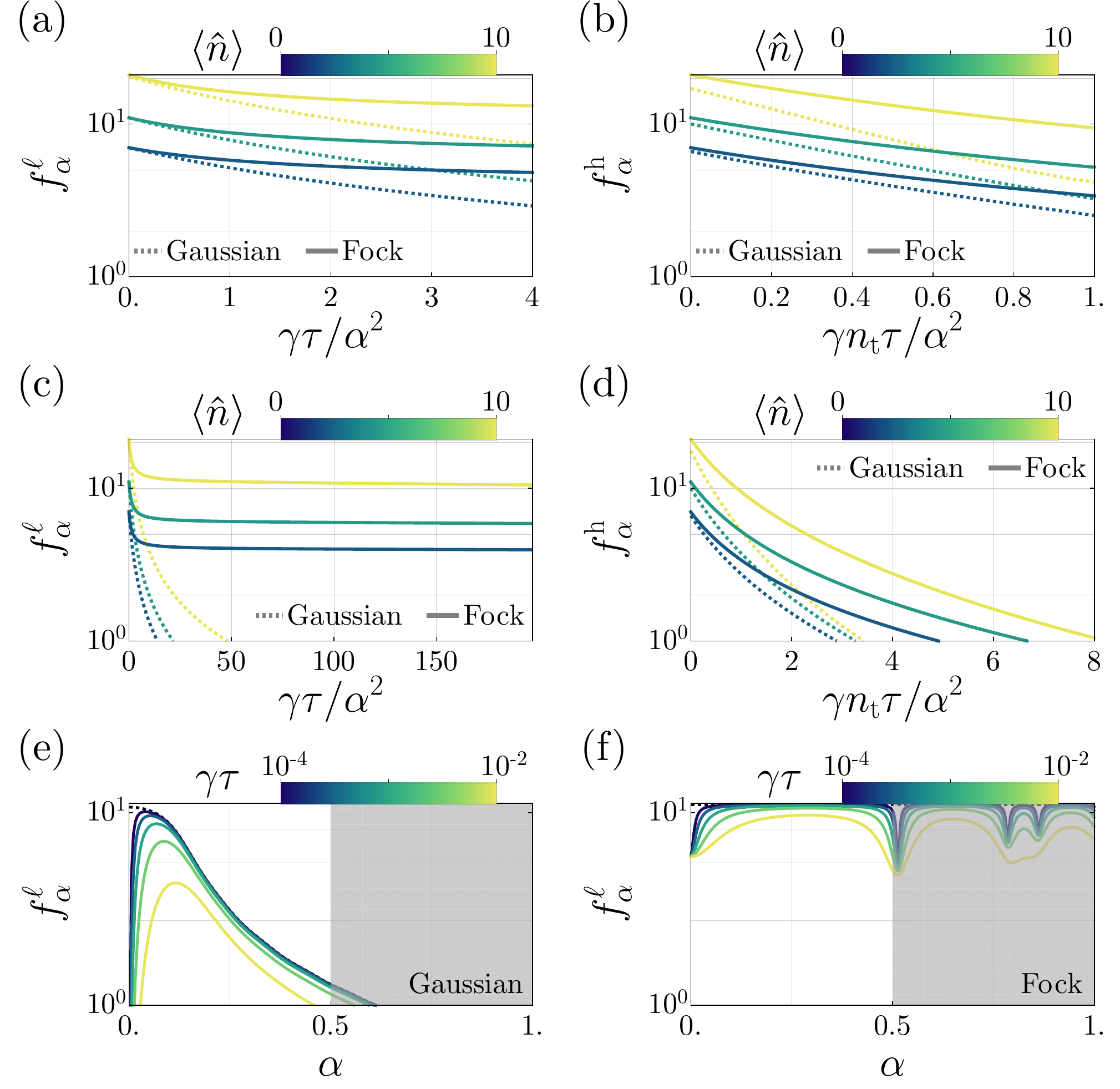}  
    \caption{Metrological gain $\fisher_{\PhaseSpaceDispAmp}$ for Fock (solid) and Gaussian (dashed) states under loss [$\ell$; (a,c,e,f)] or heating [h; (b,d)]. (a)-(d) $\fisher_{\PhaseSpaceDispAmp}$ for various $\angles{\NumberOperator}$ at fixed $\PhaseSpaceDispAmp = 0.005$ and $\PhaseSpaceDispAmp = 0.07$ [(a,c) and (b,d), respectively], computed via the approximation~\eqref{small-time} [(a,b)] or full dynamics~\eqref{full-master} [(c,d)]. Fock states outperform Gaussian ones. (e,f) $\fisher_{\PhaseSpaceDispAmp}^{\LossLetter}$ for Gaussian and Fock states with $\angles{\NumberOperator}=5$ at finite $\PhotonLossCoeff \Time$, plotted against $\PhaseSpaceDispAmp$. The shaded region marks sup-vacuum displacements $\PhaseSpaceDispAmp>0.5$. Dashed line indicates $\PhotonLossCoeff \Time = 0$. Fock states maintain gain across the whole range, unlike Gaussian ones.\label{Fig2}}
\end{figure}

\textit{Noisy state preparation.}---Having analyzed the sensing step, we now turn to state preparation.
The ability to prepare high-quality states underpins the achievable metrological advantage.
However, it strongly depends on the limitations of the physical system—especially coherence time and controllability, determined by, e.g., bandwidth for time-dependent control and strength of nonlinearity~\cite{Rosiek2024}.
To fully leverage available resources and mitigate decoherence, quantum optimal control techniques have become essential in practical state preparation~\cite{Glaser2015, Koch2022}.
As a relevant example, we consider a minimal spin-boson system with a time-dependent single-excitation spin drive and Jaynes–Cummings coupling~\cite{Rahman2025},
\begin{align}
    \hamiltonian_{\SpinBosonLetter}  (\Time) = \sigmaDownOp \PhoCre + \RenOmegaDrive (\Time) \sigmaUpOp + \text{H.c.},
    \label{spin-boson}
\end{align}
where the Hamiltonian $\hamiltonian_{\SpinBosonLetter} (\Time)$ is written in a double rotating frame with respect to both spin and bosonic degrees of freedom, $\sigmaUpOp$ creates an excitation in a two-level subsystem, and $\RenOmegaDrive (\Time)$ is a complex qubit drive.
This Hamiltonian~\cite{Liu2024} can be realized in various cavity- and circuit-QED platforms~\cite{Heeres2017,Krisnanda2025,Pan2025}, as well as with trapped ions~\cite{Matsos2024,Valahu2025} and quantum acoustodynamical systems~\cite{Rahman2025}.
Again, we consider lossy dynamics governed by~\eqref{full-master}.
For control optimization, we employ the dressed chopped random-basis technique (dCRAB)~\cite{Muller2022}.
Here, the control functions $\PosControl(\Time) \in \curlies{\text{Re}\ \RenOmegaDrive (\Time),\text{Im}\ \RenOmegaDrive (\Time)}$ are expanded into a Fourier basis with randomized frequencies and a high-frequency cutoff, approximating experimentally accessible bandwidths, $\PosControl(\Time) = \ScalingFunction(\Time) \sum_{k=1}^{N_{\text{p}}} a_k \cos \nu_k \Time + b_k \sin \nu_k \Time$, where $\ScalingFunction(\Time)$ is a scaling function, $a_k$ and $b_k$ are parameters to be optimized, $\nu_k$ are frequencies randomly drawn from a uniform distribution in $[0, \nu_\text{max}]$, $\nu_\text{max}$ is the frequency cutoff, and $N_{\text{p}}$ is the number of components.
We choose this gradient-free method as it accommodates multiple reward functions and naturally implements a frequency cutoff.

Typically, a fidelity-based functional is used to optimize FI, targeting a specific \textit{a priori} chosen state~\cite{Matsos2024, Rahman2025}.
It has also been proposed to optimize directly for metrological gain~\cite{Liu2017,Liu2017a,Pang2017,Basilewitsch2020,Lin2022}.
Here, we develop a tailored, easily computable FI-based functional that also accounts for decoherence susceptibility and the analytical bound~\eqref{bound}.
For the decoherence part, using the recipe for states under small decoherence via~\eqref{small-time}, we introduce $\Functional_{\LossLetter/\HeatLetter} \squares{\DensityMatrix; \PhotonLossCoeff \Time, \PhaseSpaceDispAmp } = \fisher_{\PhaseSpaceDispAmp}^{\LossLetter/\HeatLetter} \squares{\DensityMatrix; \PhotonLossCoeff \Time, \PhaseSpaceDispAmp }$.
To capture FI scaling with $\angles{\NumberOperator}$, we define $ \FunNumber\squares{\DensityMatrix; \AvN } = (\angles{\NumberOperator}_{\DensityMatrix}^2 / \AvN^2 )\Heaviside (\AvN - \angles{\NumberOperator}_{\DensityMatrix})$, which drives optimization toward higher occupations without affecting sensitivity once $\angles{\NumberOperator}_{\DensityMatrix} > \AvN$.
Here $\Heaviside (\cdot)$ is the Heaviside step function.
We then maximize the combined functional
\begin{align}
    \Functional_\lag \squares{\DensityMatrix; \PhotonLossCoeff \Time, \AvN ,\PhaseSpaceDispAmp }= \FunLoss \squares{\DensityMatrix; \PhotonLossCoeff \Time, \PhaseSpaceDispAmp } + \lag \FunNumber\squares{\DensityMatrix; \AvN },
\end{align}
with $\TimeMax$, $\PhaseSpaceDispAmp$, and $\lag=10$ fixed, under coherent dynamics.
In the examples, we use $\FunLoss$, though $\FunHeat$ or a combination can be chosen depending on the setup.
$\PhotonLossCoeff \Time$ is taken nonzero, ensuring operation beyond the regime~\eqref{perturbative}.
$\AvN$ is set so that $\angles{\NumberOperator}_{\DensityMatrix}>\AvN$, focusing optimization on higher $\fisher_{\PhaseSpaceDispAmp}$.
\begin{figure}[t!]
    \includegraphics[width=\linewidth]{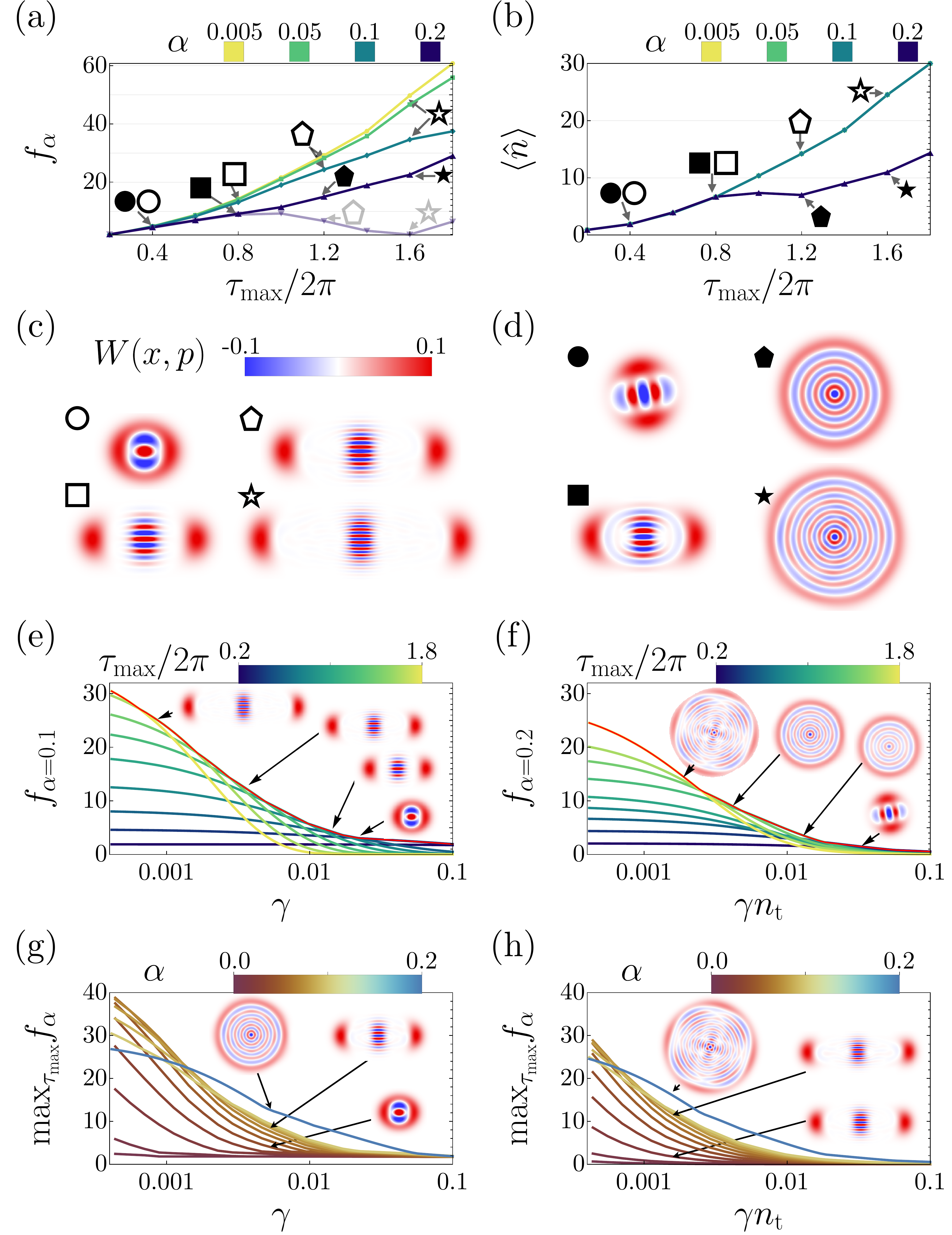}  
    \caption{(a) Metrological gain $\fisher_{\PhaseSpaceDispAmp}$ in a spin-boson system maximized via an optimal drive. The solutions for $\PhaseSpaceDispAmp \leq 0.1$ are identical for a fixed $\TimeMax$. The low-opacity curve corresponds to the states optimized for $\PhaseSpaceDispAmp = 0.005$, but evaluated for $\PhaseSpaceDispAmp = 0.2$. (b) Corresponding occupation numbers $\angles{\NumberOperator}$. The transition for $\PhaseSpaceDispAmp = 0.2$ is visible. (c,d) Wigner functions of the optimal solutions for $\PhaseSpaceDispAmp = 0.005, 0.05, 0.1$ (c) and $\PhaseSpaceDispAmp = 0.2$ (d). (e,f) Metrological gain $\fisher_{\PhaseSpaceDispAmp}$ at the end of the state preparation for the optimal states under loss with $\PhaseSpaceDispAmp = 0.1$ (e) and heating with $\PhaseSpaceDispAmp = 0.2$ (f). Red lines signify the maximum achievable FI at a fixed decoherence strength. (g,h) Maximally achievable FI for various values of $\PhaseSpaceDispAmp$. Curves are constructed analogously to the red lines from (e,f). At lower decoherence levels, consistent with (a), optimal solutions yield higher FI for smaller $\PhaseSpaceDispAmp$, with the trend reversing at larger values of $\PhotonLossCoeff$.}
 \label{Fig3}
\end{figure}

The optimization results are summarized in Fig.~\ref{Fig3}.
For unitary state preparation, Fig.~\ref{Fig3}(a) shows that $\fisher_{\PhaseSpaceDispAmp}$ grows monotonically with the total state preparation time $\TimeMax$ and decreases with $\PhaseSpaceDispAmp$.
The occupation number [Fig.~\ref{Fig3}(b)] behaves qualitatively differently for $\PhaseSpaceDispAmp \leq 0.1$ and $\PhaseSpaceDispAmp=0.2$, signaling a metrological transition~\cite{Beato2024} (understood as qualitative change of the optimal probe state as a function of the relevant parameters), further illustrated via Wigner functions in Figs.~\ref{Fig3}(c,d).
For $\PhaseSpaceDispAmp \leq 0.1$, the optimal probes are empirically well approximated by number-squeezed cat (moon) states~\cite{deMatosFilho1996,Dodonov2002,Rojkov2024,Rousseau2025}, with number distribution
\begin{align}
\DensityMatrixElement_n \sim \exp[-\FockScale \tfrac{(n - \NumberCat)^2}{2 \NumberCat}] \tfrac{\squares{\CohDis^n \pm \rounds{-\CohDis}^n}^2}{n!},
\label{moon}
\end{align}
where $\pm$ sets parity, $\NumberCat=\CohDis^2\tanh{\CohDis^2}$, and $\CohDis$ sets state size.
These interpolate between cat states ($\FockScale\to0$) and Fock states ($\FockScale\to\infty$), and consistently outperform Gaussian states under the same amount of decoherence (cf. SM~\cite{SuppMat4}).
Their optimality is also supported by Eq.~\eqref{parity-scaling}, which shows that for nonzero $\PhaseSpaceDispAmp$, states with reduced variance $\Delta^2\NumberOperator$ perform better.
At $\PhaseSpaceDispAmp=0.2$, moon states remain optimal for $\TimeMax/2\pi \leq 0.8$, but for longer times the optimization yields either Fock states or 4-spaced mixtures (e.g., $\ket{11}+\ket{15}$ at $\TimeMax/2\pi=1.6$).
This indicates that FI’s dependence on $\PhaseSpaceDispAmp$ becomes important: at larger values, more widely spaced states are favored, with deviations from the bound~\eqref{bound} scaling as $\mathcal{O}(\PhaseSpaceDispAmp^4)$ [cf. Eq.~\eqref{N-scaling}].
The emergence of moon rather than Fock states in the optimization further suggests they are easier to prepare with the same resources while considering their sensitivity.
Only at sufficiently large $\PhaseSpaceDispAmp$ do Fock states provide a practical performance and preparation advantage.

The presented solutions protect against decoherence during sensing.
To assess loss and heating during the state preparation step, we analyze optimal coherent solutions under noisy evolution, governed by the same Eq.~\eqref{full-master} as in the sensing step.
The optimal $\TimeMax$ depends on decoherence strength: stronger noise penalizes highly excited states that take longer to generate.
Figs.~\ref{Fig3}(e–h) show the maximally achievable metrological gain versus loss or heating for different $\PhaseSpaceDispAmp$.
There, the metrological transition is also visible, as the spacedness of optimal probe states varies as a function of decoherence strength.

\textit{Conclusions.}---To conclude, we introduced new families of metrologically advantageous states for phase-insensitive force sensing and analyzed their susceptibility to decoherence, showing that Fock states are generally the most robust and offer the largest dynamical range.
We also developed a noise-dependent reward functional, maximized via quantum optimal control in a spin-boson system, and found that either number-squeezed cat ~\cite{deMatosFilho1996,Dodonov2002,Rojkov2024,Rousseau2025} or Fock states yield the best sensitivities under realistic constraints.
Our framework separates the sensing protocol into preparation, sensing, and measurement stages, each subject to distinct trade-offs determined by decoherence rates, interaction strengths, and displacement mechanisms~\cite{Rahman2025}.
While the leading decoherence sources are platform-dependent, reported implementations of optimal control in bulk acoustic resonators~\cite{Rahman2025}, circuit QED platforms~\cite{Heeres2017,Pan2025}, and trapped ions~\cite{Matsos2024} imply mechanical decoherence coefficients of order $\PhotonLossCoeff \sim 10^{-3}$, $10^{-5}\!-\!10^{-4}$, and $10^{-5}$, respectively~\footnote{We take $\PhotonLossCoeff=1/g T$. The reported values read: $\JCCoup=2\pi\cdot\{2.2,1.4,0.29,0.002\}$ MHz and $T=\{2.7,1,0.089,5000\}$ ms for~\cite{Pan2025,Heeres2017,Rahman2025,Matsos2024}, respectively.}.
In this regime, our results predict at least several-dozen-fold improvements in asymptotic single-shot FI at specific coupling strengths across all these platforms, although other decoherence channels, such as qubit relaxation or motional dephasing, can impose stricter limits.
These considerations highlight broad applicability to mechanical~\cite{Matsos2024,Valahu2025,Rahman2025}, optical~\cite{Deffner2014}, and microwave~\cite{Heeres2017,Krisnanda2025,Pan2025} systems.
While our analysis focused on a spin-boson model, it naturally extends to other platforms, especially those with weak nonlinearities~\cite{Millen2020,Gonzalez-Ballestero2021,Grochowski2025a,Roda-Llordes2024b,Casulleras2024,Rosiek2024}.
Exploring transitions across broader metrological landscapes remains an open direction for future work.

\begin{acknowledgments}
\textit{Acknowledgments.}---We thank Matteo Fadel, Zdeněk Hradil, and Kimin Park for useful discussions and comments.
P.T.G. was supported by the project CZ.02.01.01/00/22\_008/0004649 (QUEENTEC) of EU and project 23-06308S of the Czech Science Foundation.
R.F. was supported by the project 21-13265X of the Czech Science Foundation.
We acknowledge open-source packages used for numerical simulations and optimization, QuTiP~\cite{Johansson2012,Johansson2013,Lambert2024} and Quantum Optimal Control Suite~\cite{Rossignolo2023}, respectively.
Data analysis and simulation codes are available on Zenodo~\cite{zenodo2}.
Some of the plots have used Scientific colour maps~\cite{Crameri2020,Crameri2023}.
\end{acknowledgments}

\bibliography{ForceSensing}


\onecolumngrid

\vspace*{5px}
\begin{center}\textbf{ \large End Matter} \end{center}
\vspace*{5px}
\twocolumngrid

\renewcommand{\theequation}{A\arabic{equation}}
\setcounter{equation}{0}

\renewcommand{\thefigure}{A\arabic{figure}}
\setcounter{figure}{0}

\phantomsection

\label{appA}
\textit{Appendix A: Derivation of Eq.~\eqref{N-scaling}.}---The starting point is the expansion of Eq.~\eqref{probneq} around $\PhaseSpaceDispAmp = 0$,
\begin{align}
\Prob_m^{\DispLetter} = \sum_{k=0}^{\infty} \DensityMatrixElement_k \sum_{r=0}^{\infty} \delta_{k,m \pm r} \FuncProbExp_{m \pm r,m},
\end{align}
where term $\FuncProbExp_{m \pm r,m} = \mathcal{O}(\PhaseSpaceDispAmp^{2r})$ includes contribution to $\Prob_m^{\DispLetter}$ involving only $\DensityMatrixElement_{m \pm r}$, i.e., the $r$-nearest \textit{neighbor}.
Here, the subscript $i$ in $\DensityMatrixElement_i$ is considered as a \textit{site} in the Fock distribution.
Note that the contribution from the nearest occupied neighbor (minimal $r$ for which $\DensityMatrixElement_{m \pm r} \neq 0$) is leading for a given site $m$ with respect to order in $\PhaseSpaceDispAmp$.
The contribution to the total metrological gain $\fisher_{\PhaseSpaceDispAmp}$ coming from the site $m$, such that $\fisher_{\PhaseSpaceDispAmp} = \sum_{m} \fisher_{m}$, depends only on $\Prob_m^{\DispLetter}$ and equals $\fisher_{m} = (\partial_{\PhaseSpaceDispAmp} \Prob_m^{\DispLetter}  )^2/ 4 \Prob_m^{\DispLetter} $.
One can then straightforwardly show that
\begin{align}
    \Prob_m^{\DispLetter} &= \DensityMatrixElement_m + \mathcal{O}(\PhaseSpaceDispAmp^2) \  &\rightarrow \ \ \ \fisher_m &= \mathcal{O}(\PhaseSpaceDispAmp^2), \label{uprow} \\
    \Prob_m^{\DispLetter} &=  \mathcal{O}(\PhaseSpaceDispAmp^{2r}) \  &\rightarrow \ \ \ \fisher_m &= \mathcal{O}(\PhaseSpaceDispAmp^{2(r-1)}), \label{downrow}
\end{align}
where \eqref{uprow} corresponds to the initially occupied Fock state $\DensityMatrixElement_m \neq 0$, while \eqref{downrow} corresponds to the unoccupied one, and it means that the leading order of $\Prob_m^{\DispLetter}$ implies the leading order of $\fisher_m$.
Let us consider $N$-spaced state, such that $\DensityMatrixElement_n \neq 0$ for a specific $n$ (see Fig.~\ref{FigA1}).
As the leading contribution in $\PhaseSpaceDispAmp$ to $\Prob_m^{\DispLetter}$ and hence to $\fisher_m$ comes from the nearest occupied neighbor, then the leading contribution to the total $\fisher_\PhaseSpaceDispAmp$ stemming from the site $n$, $\fisher_{(n)}$, equals 
\begin{align}
    \fisher_{(n)} = \sum_{r = -r_\text{m}}^{r_\text{m}} \fisher_{n+r}, \label{fishn}
\end{align}
with $r_\text{m} = \lfloor N/2 \rfloor$ is chosen such that that within range of sites $[n-r_\text{m},n+r_\text{m}]$, $\Prob_n^{\DispLetter}$ leads to the leading contribution in every site.
Here, $f_{n+N/2}$ in the even-$N$ case has contributions from both $n$- and $n+N$-sites, however, one can straightforwardly show that they linearly separate in the leading order, $f_{n+N/2} = f_{n+N/2,n} + f_{n+N/2,n+N}$, where the first term is the contribution from $n$-site, and the second from the $n+N$ one.
It follows from the equality $\FuncProbExp_{n,n+N/2} = \FuncProbExp_{n+N,n+N/2}$, which stems from the symmetry of Eq.~\eqref{probneq}.
We add these terms to $\fisher_{(n)}$ and $\fisher_{(n+N)}$, respectively.
Then, we have  $\Prob_{n+r}^{\DispLetter} = \DensityMatrixElement_n \FuncProbExp_{n,n+r} + \mathcal{O}(\PhaseSpaceDispAmp^{2(|r|+1)})$, from which follows 
\begin{align}
    \fisher_{(n)} = \frac{\DensityMatrixElement_n}{4} \sum_{r = -r_\text{m}}^{r_\text{m}} \frac{(\partial_{\PhaseSpaceDispAmp} \FuncProbExp_{n,n+r}  )^2}{\FuncProbExp_{n,n+r}} + \mathcal{O}(\PhaseSpaceDispAmp^{2 r_\text{m}}).
\end{align}
\begin{figure}[ht!]
    \includegraphics[width=\linewidth]{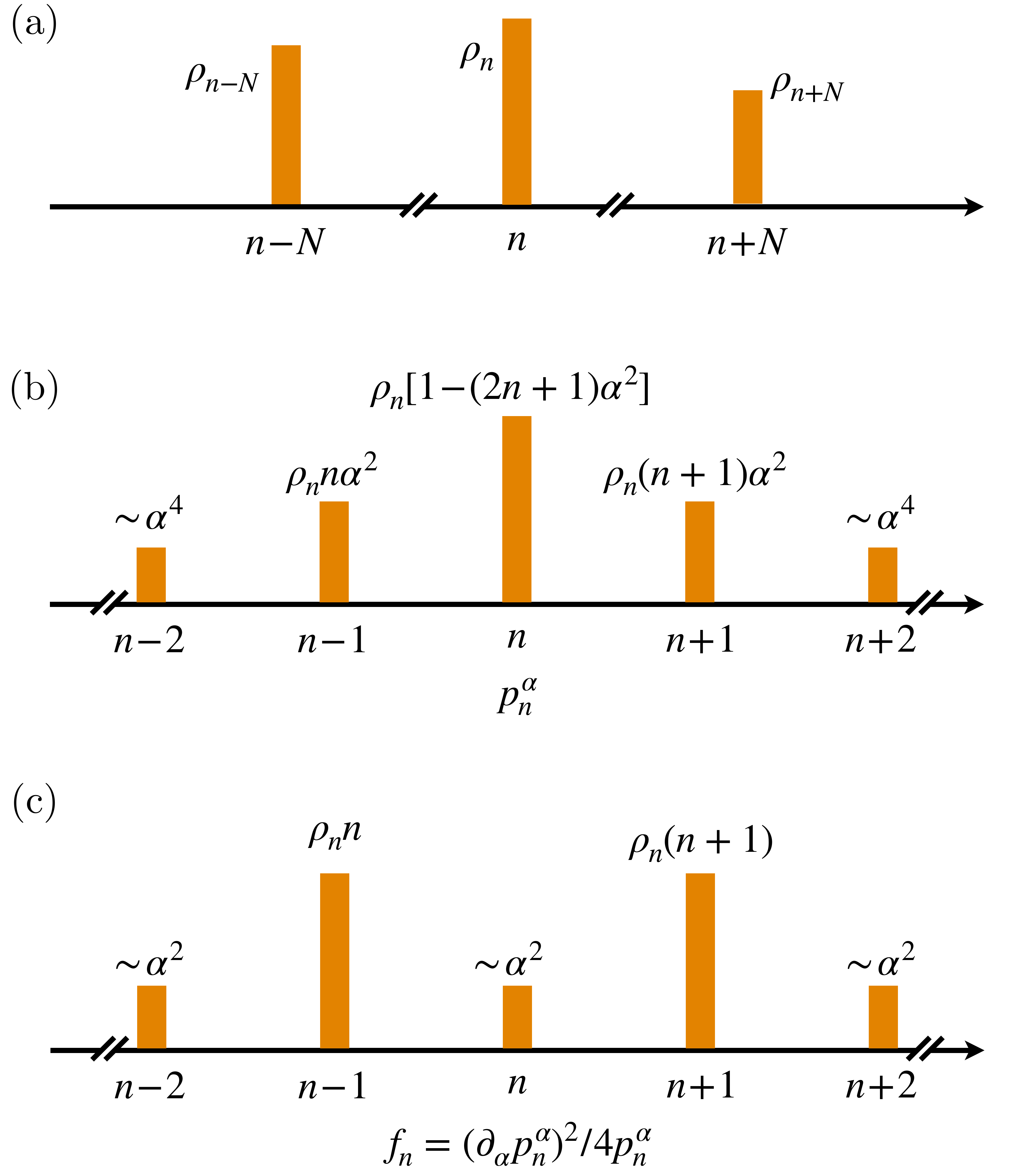}  
    \caption{(a) Number occupation of $N$-spaced state---only every $N^{\text{th}}$ Fock state is occupied. (b) Effect of the action of channel~\eqref{mixing} in the closest vicinity to some initially occupied $\DensityMatrixElement_n \neq 0$. In the limit of small $\PhaseSpaceDispAmp$, the change in probability distribution becomes smaller as the distance from the $n$-site grows. (c) Contributions to the FI from respective sites. Nonzero values come only from $n+1$- and $n$-sites.
 \label{FigA1}}
\end{figure}
We know that~\cite{Oh2020,Ritboon2022}
\begin{align}
    \DensityMatrixElement_n (1+2n) &=  \DensityMatrixElement_n  \fisher_\PhaseSpaceDispAmp (\ket{n}) = \frac{\DensityMatrixElement_n }{4} \sum_{r = -\infty}^{\infty} \frac{(\partial_{\PhaseSpaceDispAmp} \FuncProbExp_{n,n+r}  )^2}{\FuncProbExp_{n,n+r}} \nonumber \\
    & = \fisher_{(n)} + \mathcal{O}(\PhaseSpaceDispAmp^{2 \lfloor N/2 \rfloor }) + \frac{\DensityMatrixElement_n }{4} \sum_{|r| > r_\text{m}} \frac{(\partial_{\PhaseSpaceDispAmp} \FuncProbExp_{n,n+r}  )^2}{\FuncProbExp_{n,n+r}}, 
\end{align}
so $\fisher_{(n)} = \DensityMatrixElement_n (1+2n) - \mathcal{O}(\PhaseSpaceDispAmp^{2 \lfloor N/2 \rfloor }) $, and
\begin{align}
   \fisher_\PhaseSpaceDispAmp = \sum_n \fisher_{(n)}  = 1 + 2 \angles{\NumberOperator}  - \mathcal{O}(\PhaseSpaceDispAmp^{2\lfloor N/2 \rfloor} ).
\end{align}
Note that the leading contributions to $\fisher_{(n)}$ come from $n-1$ and $n+1$ sites and read $\fisher_{n-1} = \DensityMatrixElement_n n + \mathcal{O} (\PhaseSpaceDispAmp^2)$ and $\fisher_{n+1} = \DensityMatrixElement_n (n+1) + \mathcal{O} (\PhaseSpaceDispAmp^2)$ [cf. Fig.~\ref{FigA1}(c)].
It is also clear why 1-spaced states do not offer an advantage---each contribution to $\fisher_\PhaseSpaceDispAmp$ is of the order $\mathcal{O}(\PhaseSpaceDispAmp^2)$ as follows from Eq.~\eqref{uprow}.

\renewcommand{\theequation}{B\arabic{equation}}
\setcounter{equation}{0}

\renewcommand{\thefigure}{B\arabic{figure}}
\setcounter{figure}{0}

\textit{Appendix B: Derivation of Eq.~\eqref{parity-scaling}\label{appB}.}---Let us consider a $2$-spaced state, where $\DensityMatrixElement_n \neq 0$.
Then, expanding Eq.~\eqref{probneq}, following the discussion in App.~\hyperref[appA]{A}, the relevant occupations read
\begin{align}
    \Prob_n^{\DispLetter}  =& \DensityMatrixElement_n - \DensityMatrixElement_n (2n+1) \PhaseSpaceDispAmp^2 + \mathcal{O} (\PhaseSpaceDispAmp^4), \nonumber \\
    \Prob_{n+1}^{\DispLetter}  =& \DensityMatrixElement_n \squares{ \rounds{n+1} \PhaseSpaceDispAmp^2 - \rounds{n+1}^2 \PhaseSpaceDispAmp^4} + \nonumber \\ &\DensityMatrixElement_{n+2} \squares{ \rounds{n+2} \PhaseSpaceDispAmp^2 - \rounds{n+2}^2 \PhaseSpaceDispAmp^4} + \mathcal{O} (\PhaseSpaceDispAmp^6).
\end{align}
The contribution to the metrological gain $\fisher_\PhaseSpaceDispAmp$ from these two terms reads
\begin{align}
    &\bar{\fisher}_n =\fisher_n + \fisher_{n+1} = \rounds{n+1} \DensityMatrixElement_n + \rounds{n+2} \DensityMatrixElement_{n+1} + \nonumber \\
    & \PhaseSpaceDispAmp^2 \squares{ \rounds{n^2-2n-2} \DensityMatrixElement_n - 3 \rounds{n+2}^2 \DensityMatrixElement_{n+2}} + \mathcal{O}(\PhaseSpaceDispAmp^4),
\end{align}
which, after summation, amounts to
\begin{align}
    \fisher_\PhaseSpaceDispAmp = \sum_n  &\bar{\fisher}_n  = 1 + 2 \angles{\NumberOperator} - 2 \PhaseSpaceDispAmp^2 \rounds{1 +  \angles{\NumberOperator} + \angles{\NumberOperator^2}} + \mathcal{O}(\PhaseSpaceDispAmp^4 ) .
\end{align}

\renewcommand{\theequation}{C\arabic{equation}}
\setcounter{equation}{0}

\renewcommand{\thefigure}{C\arabic{figure}}
\setcounter{figure}{0}

\textit{Appendix C: Comparison of FI for different quantum non-Gaussian states\label{appC}.}---In the main text, we have introduced Fock, Gaussian [Eq.~\eqref{gaussian}], and moon [Eq.~\eqref{moon}] states.
Beyond them, we also analyze other typical metrologically useful families of 2- and 4-spaced states.
First, finite-energy grid (GKP) states, whose number distribution is numerically evaluated from their spatial representation,
\begin{align}
    \psi_{\text{GKP}} (x,\Delta)= \Normalization_{\Delta} \sum_{k=-\infty}^{\infty} e^{-k^2 \pi \Delta^2 } \frac{1}{\sqrt{\Delta \sqrt{\pi}}} e^{-\frac{(x-\sqrt{2 \pi}k)^2}{2 \Delta^2}},
\end{align}
where $\Normalization_{\Delta} \approx 1/\sqrt{\vartheta_3\rounds{0,e^{-2 \pi \Delta^2}}}$.
Here, $\vartheta_3$ is Jacobi elliptic theta function.
Finite-energy grid states are 2-spaced as they are mirror-symmetric in $x$ and they are approximately 4-spaced when $\Delta \rightarrow 0$.
Then, the compass states, for which we have
\begin{align}
    \DensityMatrixElement_n = \Normalization_{\CohDis}   \frac{\left|\CohDis^n + \rounds{-\CohDis}^n + \rounds{i\CohDis}^n + \rounds{-i\CohDis}^n \right|^2}{n!},
    \label{compass}
\end{align}
with $\Normalization_{\CohDis} = 1/\sqrt{8 \rounds{\cos \CohDis^2 + \cosh \CohDis^2}}$.
They are explicitly 4-spaced.
The last family is the number-phase states,
\begin{align}
\ket{\WaveFunction}_{\text{np}} = \Normalization_{\mu}\sum_{k=0}^{\infty} \text{Ai} \squares{\rounds{\frac{\mu}{N^2}}^{1/3} N (k+1) - |z_1|} \ket{kN+q}
\end{align}
where $\Normalization_{\mu,N}$ is the numerically evaluated normalization constant, $\mu$ sets the average occupation number, $|z_1| \approx 2.338$ is the absolute value of the first zero of the Airy function $\text{Ai}(x)$~\cite{Valahu2025}, and $q$ is the offset.
These states are explicitly $N$-spaced.
In Fig.~\ref{FigC1} we compare FI for these states.

\begin{figure}[ht!]
    \includegraphics[width=\linewidth]{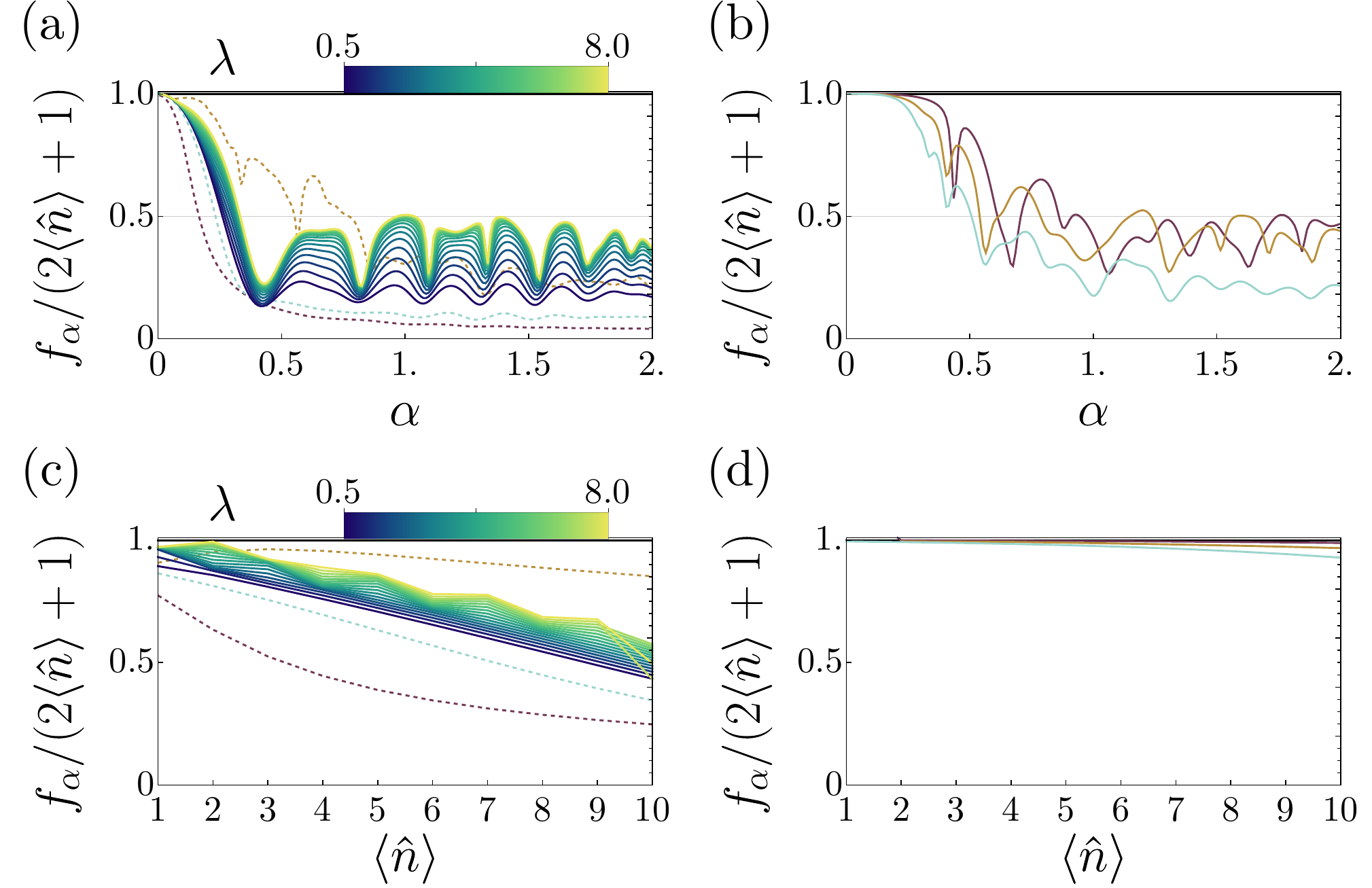}  
    \caption{Metrological gain $\fisher_\PhaseSpaceDispAmp$ for different families of non-Gaussian states as a function of $\PhaseSpaceDispAmp$ (a,b) and $\angles{\NumberOperator}$ (c,d). The families include 2-spaced states (a,c): moon (green-to-blue solid lines), Gaussian (dashed purple), GKP (dashed brown), number-phase (dashed sky blue), and 4-spaced states (b,d): $\ket{3}+\ket{7}$ (solid purple), compass (solid brown), and number-phase (solid blue sky) states. Solid black lines represent Fock state. Either $\langle \NumberOperator \rangle=5$ is fixed (a,b) or $\PhaseSpaceDispAmp=0.05$ (c,d). The considered non-Gaussian states outperform the Gaussian ones, with 4-spaced states performing almost as well as Fock states for small values of $\PhaseSpaceDispAmp$ and over a broad range of $\langle \NumberOperator \rangle$.}
 \label{FigC1}
\end{figure}

\renewcommand{\theequation}{D\arabic{equation}}
\setcounter{equation}{0}

\renewcommand{\thefigure}{D\arabic{figure}}
\setcounter{figure}{0}

\textit{Appendix D: Derivation of Eq.~\eqref{perturbative}\label{appD}.}---The diagonal elements of the slightly decohered state~\eqref{small-time} read (for details see SM~\cite{SuppMat4}):
\begin{align}
\DensityMatrixElement^\LossLetter_n &=
\DensityMatrixElement_n + \SmallTime\squares{  - n \DensityMatrixElement_n + \rounds{n+1} \DensityMatrixElement_{n+1}} , \nonumber \\
\DensityMatrixElement^\HeatLetter_n &= \DensityMatrixElement_n + \SmallTimeBar\squares{ n \DensityMatrixElement_{n-1} - \rounds{2n+1}\DensityMatrixElement_n + \rounds{n+1} \DensityMatrixElement_{n+1}}.
\label{ElementsDecoherenceEnd}
\end{align}
Let us compute the leading contribution due to $\SmallTime$ and $\SmallTimeBar$ in the loss and heating cases, respectively.
As follows from App.~\hyperref[appA]{A}, for at least 3-spaced states, we can focus on the leading contributions to $\fisher_n^{\LossLetter / \HeatLetter}$, namely, from $n$- and $n+1$-sites:
\begin{align}
\Prob_{n-1}^{\DispLetter, \LossLetter} &= \Prob_{n-1}^{\DispLetter}, \nonumber \\
\Prob_{n+1}^{\DispLetter, \LossLetter} &= \DensityMatrixElement_n \squares{ (1- n \SmallTime)  n \PhaseSpaceDispAmp^2  + \SmallTime n }, \nonumber \\
\Prob_{n-1}^{\DispLetter, \HeatLetter} &= \DensityMatrixElement_n \curlies{ \squares{1- \rounds{2n+1} \SmallTimeBar} \rounds{ n} \PhaseSpaceDispAmp^2  + \SmallTimeBar n }, \nonumber \\
\Prob_{n+1}^{\DispLetter, \HeatLetter} &= \DensityMatrixElement_n \curlies{ \squares{1- \rounds{2n+1} \SmallTimeBar} \rounds{ n+1} \PhaseSpaceDispAmp^2  + \SmallTimeBar \rounds{ n+1}  },
\label{ElementsDecoherenceEnd2}
\end{align}
where the expressions are exact up to $\mathcal{O}(\PhaseSpaceDispAmp^4)$ contributions.
Then, expanding in $n \SmallTime$ and $n \SmallTimeBar$, respectively, and summing, $\fisher_\PhaseSpaceDispAmp^{\LossLetter / \HeatLetter} = \sum_n\fisher_n^{\LossLetter / \HeatLetter}$, we arrive at
\begin{align}
\fisher_\PhaseSpaceDispAmp^{\LossLetter} &= \angles{\NumberOperator} +1 + \frac{\angles{\NumberOperator}}{1+\frac{\SmallTime}{\PhaseSpaceDispAmp^2}} +  \mathcal{O}(\SmallTime \PhaseSpaceDispAmp^2), \label{3spacedexp1} \\
\fisher_\PhaseSpaceDispAmp^{\HeatLetter} &= \frac{1+2\angles{\NumberOperator}}{1+\frac{\SmallTimeBar}{\PhaseSpaceDispAmp^2}} + \mathcal{O}(\SmallTimeBar \PhaseSpaceDispAmp^2). \label{3spacedexp2}
\end{align}
For 2-spaced states, the leading contributions come from $n+1$-sites that read
\begin{align}
\Prob_{n+1}^{\DispLetter, \LossLetter} &= \DensityMatrixElement_n \rounds{n+1} \PhaseSpaceDispAmp^2 +  \DensityMatrixElement_{n+2} (n+2)   \rounds{ \PhaseSpaceDispAmp^2  + \SmallTime  } + \mathcal{O}(\SmallTime \PhaseSpaceDispAmp^2), \label{2spacedexp1}\\
\Prob_{n+1}^{\DispLetter, \HeatLetter} &=\squares{ \DensityMatrixElement_n \rounds{n+1} + \DensityMatrixElement_{n+2} \rounds{n+2}  } \rounds{ \PhaseSpaceDispAmp^2  + \SmallTimeBar  } + \mathcal{O}(\SmallTimeBar \PhaseSpaceDispAmp^2).
\label{2spacedexp2}
\end{align}
Using~\eqref{2spacedexp2} leads directly to~\eqref{3spacedexp2}, while~\eqref{2spacedexp1} gives
\begin{align}
    \fisher_\PhaseSpaceDispAmp = \sum_{n} \frac{\DensityMatrixElement_{n+2} (n+2)+\DensityMatrixElement_{n} (n+1) }{1+\frac{\SmallTime}{\PhaseSpaceDispAmp^2} \frac{1}{1+\frac{\DensityMatrixElement_n (n+1)}{\DensityMatrixElement_{n+2} (n+2)}}} + \mathcal{O}(\SmallTime \PhaseSpaceDispAmp^2), 
\end{align}
which reduces to $\fisher_\PhaseSpaceDispAmp = \mathcal{O}(\PhaseSpaceDispAmp^2 / \SmallTime )$ and $\fisher_\PhaseSpaceDispAmp = 2 \angles{\NumberOperator}+1 - \SmallTime \angles{\NumberOperator}/ \PhaseSpaceDispAmp^2$ in the limits $\PhaseSpaceDispAmp \rightarrow 0$ and $\SmallTime \rightarrow 0$, respectively.

\clearpage
\title{Optimal phase-insensitive force sensing with non-Gaussian states: Supplemental Materials}
\maketitle

\makeatletter
\renewcommand{\c@secnumdepth}{0}
\makeatother

\renewcommand{\thesection}{S\Roman{section}}
\renewcommand{\thesubsection}{S\thesection.\arabic{subsection}}
\renewcommand{\thesubsubsection}{S\thesubsection.\arabic{subsubsection}}

\renewcommand{\thefigure}{S\arabic{figure}}
\renewcommand{\theequation}{S\arabic{equation}}

\makeatletter
\renewcommand{\p@subsection}{}
\renewcommand{\p@subsubsection}{}
\renewcommand{\p@figure}{}
\makeatother

\setcounter{section}{0}
\setcounter{equation}{0}
\setcounter{figure}{0}

\section{Derivation of $\DensityMatrix_\DispLetter$ and Fisher information}
\begin{figure}[h!]
    \includegraphics[width=\linewidth]{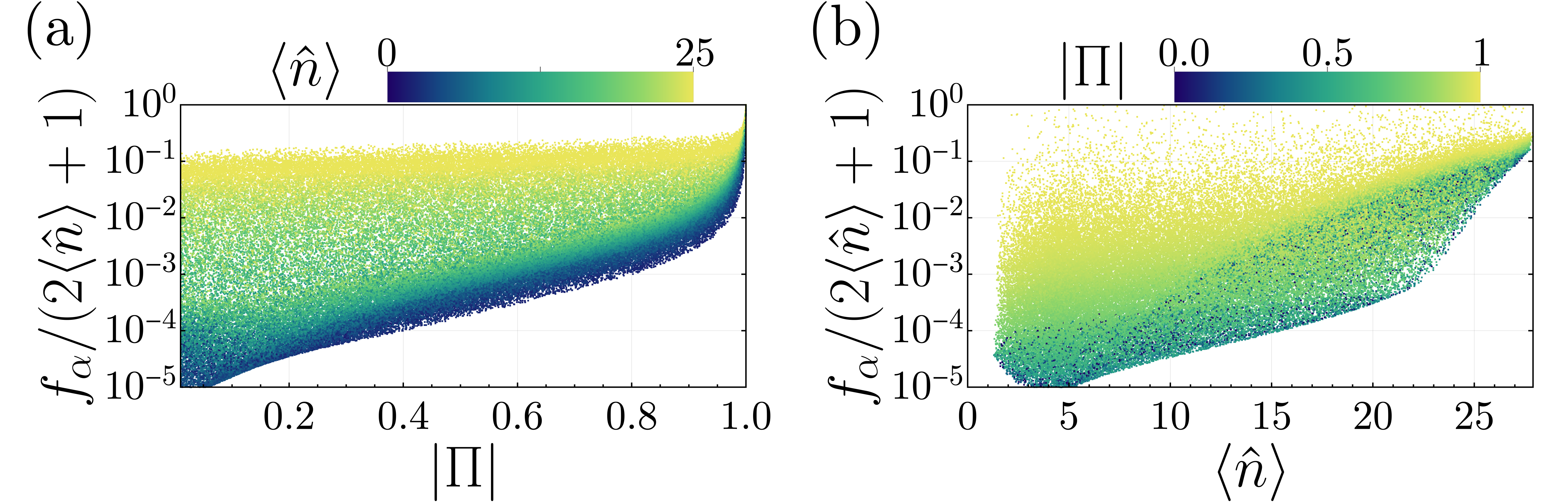}    
    \caption{Fisher information for $\PhaseSpaceDispAmp = 0.005$ for $10^5$ occupation number distributions drawn from uniform distributions in both $\Parity$ and $\angles{\NumberOperator}$ as a function of (a) parity $|\Parity|$ and (b) occupation number  $ \angles{\NumberOperator}$. These plot points correspond to dots from Fig.~\ref{FigS4}. \label{FigS1}}
\end{figure}

Here we compute the action of phase-randomization channels from the main text.
The starting point is the overlap between Fock $\ket{m}$ and displaced Fock $\ket{k,\PhaseSpaceDisp}=\Displacement{\PhaseSpaceDispAmp e^{\ImagUnit \DispPhase}} \ket{k}$~\cite{deOliveira1990},
\begin{align}
    \braket{m}{k,\PhaseSpaceDisp} =& e^{-\frac{\PhaseSpaceDispAmp^2}{2}} \rounds{\frac{m!}{k!}}^{\frac{1}{2}\text{sgn} (k-m) } \PhaseSpaceDispAmp^{|m-k|} e^{\ImagUnit \DispPhase (m-k)} \times \nonumber \\
    & L_{\text{min} (k,m)}^{|k-m|} \rounds{\PhaseSpaceDispAmp^2} (-1)^{(k-m) \Theta(k-m)} \nonumber \\
    =& \co_{km} e^{\ImagUnit \DispPhase (m-k)}.
\end{align}
Then, the channels~\eqref{channels} act as follows:
\begin{align}
    \DensityMatrix_\DispLetter^{1} &=\int_{0}^{2 \pi} \frac{\dd \DispPhase}{2 \pi} \ \Displacement{\PhaseSpaceDispAmp e^{\ImagUnit \DispPhase}} 
    \DensityMatrix \ \DisplacementDagger{\PhaseSpaceDispAmp e^{\ImagUnit \DispPhase}} \nonumber \\
    & = \sum_{klmm'} \DensityMatrixElement_{kl} \co_{km} \co_{lm'}^{*} \delta(m-k-m'+l) \ket{m}\bra{m'}, \nonumber \\
    \DensityMatrix_\DispLetter^{2} &=\int_{0}^{2 \pi} \frac{\dd \DispPhase}{2 \pi} \ \Displacement{\PhaseSpaceDispAmp} e^{-\ImagUnit \DispPhase \NumberOperator}
    \DensityMatrix \ e^{\ImagUnit \DispPhase \NumberOperator} \DisplacementDagger{\PhaseSpaceDispAmp}  \nonumber \\
    & = \sum_{klmm'} \DensityMatrixElement_{kl} \co_{km} \co_{lm'}^{*} \delta(k-l) \ket{m}\bra{m'}.
\end{align}
Then, we have
\begin{align}
    \DensityMatrix_\DispLetter^{i} &= \sum_{n=0}^{\infty} \Prob_n^{\DispLetter} \ket{n} \bra{n} +  \sum_{n \neq n'} \DensityMatrixElement_{nn'}^{i}\ket{n} \bra{n'},
\end{align}
where
\begin{align}
\Prob_n^{\DispLetter} &= \sum_{k=0}^{\infty} \DensityMatrixElement_{k} e^{-\PhaseSpaceDispAmp^2} \rounds{\frac{n!}{k!}}^{\frac{k-n}{|k-n|}} \PhaseSpaceDispAmp^{2 |k-n|} \squares{ L_{\mu}^{|k-n|} \rounds{\PhaseSpaceDispAmp^2}}^2, \nonumber \\
\DensityMatrixElement_{nn'}^{1} &= \sum_{k=\text{max}(|n-n'|,0)}^{\infty} \DensityMatrixElement_{k+n-n',k} \co_{k+n-n',k} \co_{k,n'}^{*}, \nonumber \\
\DensityMatrixElement_{nn'}^{2} &= \sum_{k=0}^{\infty} \DensityMatrixElement_{kk} \co_{kn} \co_{kn'}^{*}.
\end{align}
To compute the Fisher information,
\begin{align}
    \Fisher_{\DispLetter} = \sum_{n=0}^{\infty} \frac{1}{\Prob_n^{\DispLetter}} \rounds{\frac{\partial \Prob_n^{\DispLetter}}{\partial {\DispLetter} }}^2,
\end{align}
one can evaluate
\begin{align}
\frac{\partial \Prob_n^{\DispLetter}}{\partial {\DispLetter}} =& 2\sum_{k=0}^{\infty} \DensityMatrixElement_{k} e^{-\PhaseSpaceDispAmp^2} \rounds{\frac{n!}{k!}}^{\frac{k-n}{|k-n|}} \PhaseSpaceDispAmp^{2 |k-n|-1}  L_{\mu}^{|k-n|} \rounds{\PhaseSpaceDispAmp^2} \times \nonumber \\
&\squares{ -2 \PhaseSpaceDispAmp^2  L_{\mu-1}^{|k-n|+1} \rounds{\PhaseSpaceDispAmp^2} + \rounds{|k-n|-\PhaseSpaceDispAmp^2} L_{\mu}^{|k-n|} \rounds{\PhaseSpaceDispAmp^2}  }.
\end{align}
In the limit of small $\PhaseSpaceDispAmp$, the Fisher information can be expanded as 
\begin{align}
    &\Fisher_{\DispLetter} = 4\PhaseSpaceDispAmp^2 \Bigg[\frac{\rounds{\DensityMatrixElement_0-\DensityMatrixElement_1}^2}{\DensityMatrixElement_0} \nonumber \\
    &+  \sum_{n=1}^{\infty} \frac{\squares{ n \DensityMatrixElement_{n-1}- \rounds{2 n+1}\DensityMatrixElement_{n} + (n+1)\DensityMatrixElement_{n+1}}^2}{\DensityMatrixElement_n} \Bigg]+\mathcal{O}(\PhaseSpaceDispAmp^4 ).
    \label{falpha1}
\end{align}
In Fig.~\ref{FigS1} we plot $\fisher_{\DispLetter}=\Fisher_{\DispLetter}/4$ for random states as a function of their occupation number $ \angles{\NumberOperator}$ and parity $\Parity = \text{Tr}[\rounds{-1}^{\NumberOperator} \DensityMatrix]$.
It quantifies how close to perfectly 2-spaced a state needs to be in order to be close to the theoretical bound.

\section{Solution of the Master equation}
The Master equation for the bosonic mode coupled to the thermal bath,
\begin{align}
    \partial_{\Time} \DensityMatrix (\Time) = &- \ImagUnit \squares{\SmallFrequency \Creation \Annihilation,\DensityMatrix (\Time)} \nonumber \\
    &+ \frac{\PhotonLossCoeff \rounds{1+\ThermalOccupation}}{2} \squares{2 \Annihilation \DensityMatrix (\Time) \Creation - \DensityMatrix (\Time ) \Annihilation^{\dagger} \Annihilation - \Annihilation^{\dagger} \Annihilation \DensityMatrix (\Time )   } \nonumber \\
    &+ \frac{\PhotonLossCoeff \ThermalOccupation}{2} \squares{2 \Creation \DensityMatrix (\Time) \Annihilation - \DensityMatrix (\Time ) \Annihilation \Annihilation^{\dagger} - \Annihilation \Annihilation^{\dagger} \DensityMatrix (\Time )   }
    \label{fullHamiltonian}
\end{align}
can be explicitly solved~\cite{Fujii2013},
\begin{align}
\DensityMatrix (\Time) = &\frac{e^{\PhotonLossCoeff \Time/2}}{\FFunction\rounds{\Time}} \sum_{j=0}^{\infty} \frac{\GFunction^j\rounds{\Time}}{j!} \Annihilation^{\dagger j} e^{-\squares{\ImagUnit \SmallFrequency \Time+\log{\FFunction\rounds{\Time}}} \NumberOperator} \times \nonumber \\
& \squares{\sum_{k=0}^{\infty} \frac{\EFunction^k\rounds{\Time}}{k!} \Annihilation^{k} \DensityMatrix \Annihilation^{\dagger k}  } e^{\squares{\ImagUnit \SmallFrequency \Time - \log{\FFunction\rounds{\Time}}} \NumberOperator} \Annihilation^{ j},
\end{align}
where
\begin{align}
\FFunction\rounds{\Time} &= \cosh{\frac{\PhotonLossCoeff \Time}{2}} + \rounds{2 \ThermalOccupation +1} \sinh{\frac{\PhotonLossCoeff \Time}{2}}, \nonumber \\
\EFunction\rounds{\Time} &= \frac{2 \rounds{\ThermalOccupation+1}}{\FFunction\rounds{\Time}}\sinh{\frac{\PhotonLossCoeff \Time}{2}}, \nonumber \\
\GFunction\rounds{\Time} &= \frac{2 \ThermalOccupation}{\FFunction\rounds{\Time}}\sinh{\frac{\PhotonLossCoeff \Time}{2}}.
\end{align}
This solution can be evaluated in the limit $\PhotonLossCoeff \ThermalOccupation \Time \ll 1$, and working in the rotating frame, so that $\SmallFrequency = 0$,
\begin{align}
\DensityMatrix (\Time) \approx \DensityMatrix\rounds{1-\SmallTime \EffThermalMinus} - \SmallTime \EffThermalBar \rounds{ \DensityMatrix \NumberOperator + \NumberOperator \DensityMatrix} + \SmallTime  \EffThermalMinus \Creation \DensityMatrix \Annihilation + \SmallTime  \EffThermalPlus \Annihilation \DensityMatrix \Creation,
\end{align}
where we have introduced $\SmallTime = \PhotonLossCoeff \Time$, $\EffThermalBar = \ThermalOccupation + 1/2$, and $\EffThermal_{\pm} = \EffThermalBar \pm 1/2 $.
In the limit of $\ThermalOccupation = 0$ we recover the case of pure excitation loss,
\begin{align}
\DensityMatrix_\LossLetter =  \DensityMatrix -\frac{1}{2} \SmallTime \rounds{\DensityMatrix \NumberOperator + \NumberOperator \DensityMatrix}+ \SmallTime \Annihilation \DensityMatrix \Creation,
\end{align}
while for $\ThermalOccupation \gg 1$, we recover the heating case,
\begin{align}
\DensityMatrix_\HeatLetter =  \DensityMatrix\rounds{1-\SmallTimeBar} - \SmallTimeBar \rounds{ \DensityMatrix \NumberOperator + \NumberOperator \DensityMatrix} + \SmallTimeBar \rounds{   \Creation \DensityMatrix \Annihilation +  \Annihilation \DensityMatrix \Creation},
\end{align}
where we have introduced $\SmallTimeBar = \SmallTime \ThermalOccupation$.
At the level of diagonal elements of the density matrix, we have then
\begin{align}
\DensityMatrixElement_n& \rightarrow \DensityMatrixElement^\LossLetter_n =
\DensityMatrixElement_n + \SmallTime\squares{  - n \DensityMatrixElement_n + \rounds{n+1} \DensityMatrixElement_{n+1}} , \nonumber \\
\DensityMatrixElement_n& \rightarrow \DensityMatrixElement^\HeatLetter_n =
\nonumber \\
& \DensityMatrixElement_n + \SmallTimeBar\squares{ n \DensityMatrixElement_{n-1} - \rounds{2n+1}\DensityMatrixElement_n + \rounds{n+1} \DensityMatrixElement_{n+1}}.
\label{ElementsDecoherence}
\end{align}

\section{Different families of states}\label{supp:families}
\begin{figure}[h!]
    \includegraphics[width=\linewidth]{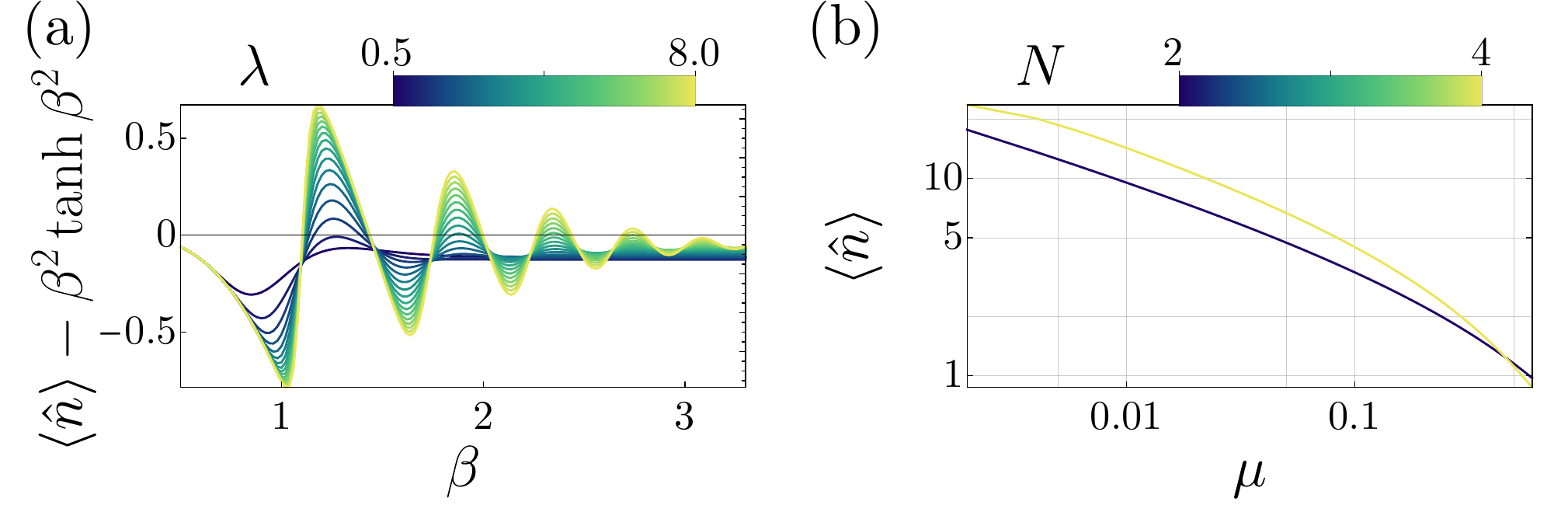}  
    \caption{Average occupation numbers as a function of relevant state parameters for the moon states (a) and number-phase states (b).   \label{FigS2}}
\end{figure}

Here we show the number distributions for the checked families of pure states.
For the Gaussian states they read:
\begin{align}
    \DensityMatrixElement_n =
    \begin{cases}
      \binom{n}{n/2} \frac{\angles{\NumberOperator}^{n/2}}{2^n \rounds{1+\angles{\NumberOperator}}^{(n+1)/2}} & \text{if $n$ is even,}\\
      0 & \text{if $n$ is odd,}
    \end{cases}
\end{align}
while for the moon states:
\begin{align}
    \DensityMatrixElement_n = \Normalization_{\FockScale,\CohDis} \exp[-\FockScale \frac{(n - \NumberCat)^2}{2 \NumberCat}]  \frac{\squares{\CohDis^n \pm \rounds{-\CohDis}^n}^2}{n!},
    \label{moonSM}
\end{align}
where $\Normalization_{\FockScale,\CohDis}$ is numerically evaluated.
The moon states are 2-spaced.
For the finite-energy grid states, the number distribution is numerically evaluated from their spatial representation,
\begin{align}
    \psi_{\text{GKP}} (x,\Delta)= \Normalization_{\Delta} \sum_{k=-\infty}^{\infty} e^{-k^2 \pi \Delta^2 } \frac{1}{\sqrt{\Delta \sqrt{\pi}}} e^{-\frac{(x-\sqrt{2 \pi}k)^2}{2 \Delta^2}},
\end{align}
where 
\begin{align}
\Normalization_{\Delta} \approx \frac{1}{\sqrt{\vartheta_3\rounds{0,e^{-2 \pi \Delta^2}}}}
\end{align}
Here, $\vartheta_3$ is Jacobi elliptic theta function, and the occupation number is approximately given by
\begin{align}
\angles{\NumberOperator} \approx \sinh^2({\log{\Delta}}) + \frac{2 \pi \sum_{k=1}^{\infty} k^2 e^{-2 k^2 \pi \Delta^2}}{\vartheta_3\rounds{0,e^{-2 \pi \Delta^2}}}.
\end{align}
Finite-energy grid states are 2-spaced as they are mirror-symmetric in $x$ and they are approximately 4-spaced when $\Delta \rightarrow 0$.
For the compass states, we have
\begin{align}
    \DensityMatrixElement_n = \Normalization_{\CohDis}   \frac{\left|\CohDis^n + \rounds{-\CohDis}^n + \rounds{i\CohDis}^n + \rounds{-i\CohDis}^n \right|^2}{n!},
    \label{compassSM}
\end{align}
with
\begin{align}
 \Normalization_{\CohDis} = \frac{1}{\sqrt{8 \rounds{\cos \CohDis^2 + \cosh \CohDis^2}}},
\end{align}
and 
\begin{align}
\angles{\NumberOperator} = \frac{\CohDis^2\rounds{-\sin \CohDis^2 + \sinh \CohDis^2}}{\cos \CohDis^2 + \cosh \CohDis^2}    
\end{align}
They are explicitly 4-spaced.
The last family is the number-phase states,
\begin{align}
\ket{\WaveFunction}_{\text{np}} = \Normalization_{\mu}\sum_{k=0}^{\infty} \text{Ai} \squares{\rounds{\frac{\mu}{N^2}}^{1/3} N (k+1) - |z_1|} \ket{kN+q}
\end{align}
where $\Normalization_{\mu,N}$ is the normalization constant, $\mu$ sets the average occupation number, $|z_1| \approx 2.338$ is the absolute value of the first zero of the Airy function $\text{Ai}(x)$~\cite{Valahu2025}, and $q$ is the offset.
These states are explicitly $N$-spaced.
For completeness, in Fig.~\ref{FigS2} we plot the average occupation numbers for the moon and number-phase states as a function of $\FockScale$ and $\CohDis$, and $\mu$, respectively.

\begin{figure*}[ht!]
    \includegraphics[width=\linewidth]{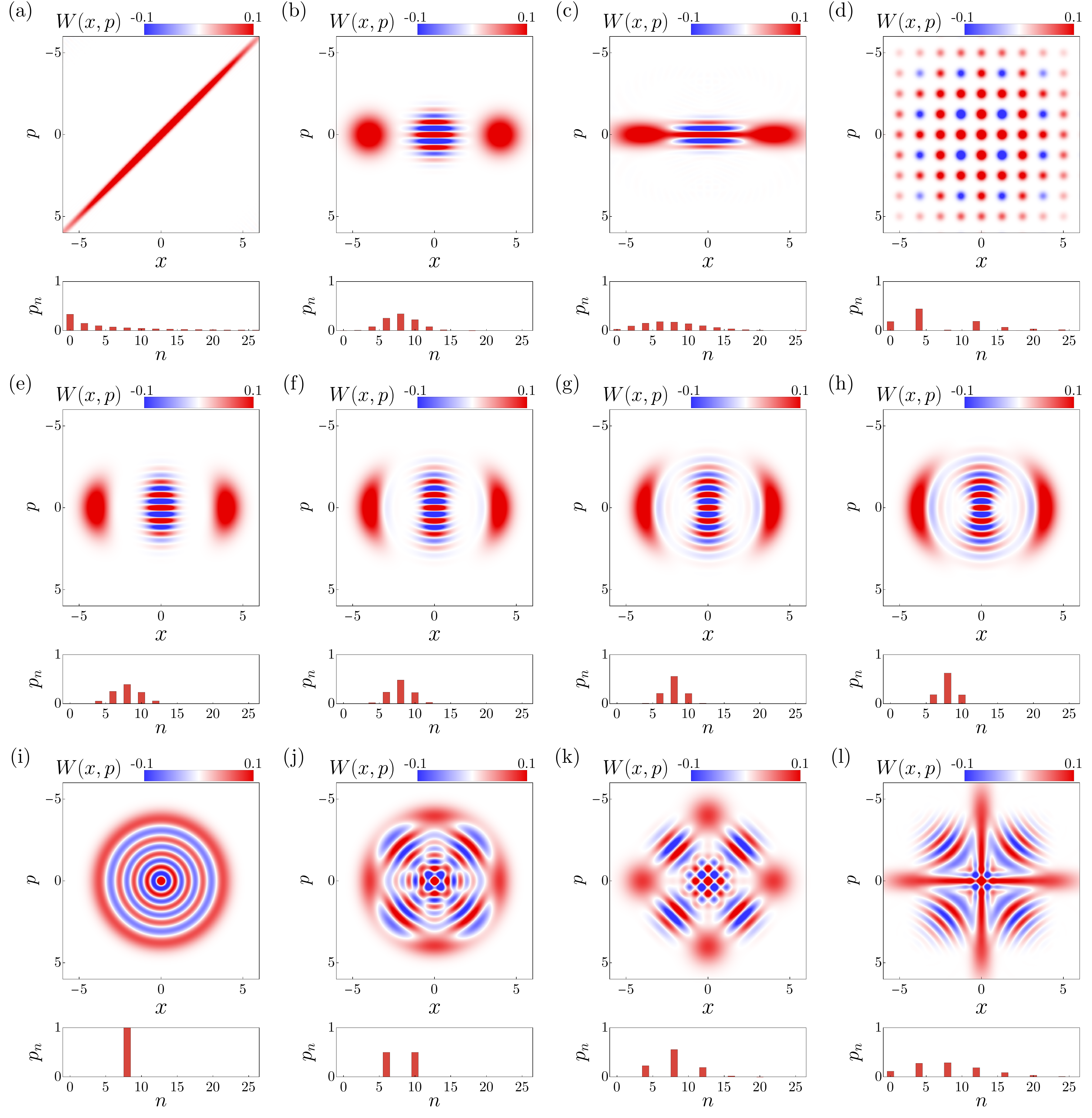}  
    \caption{The zoo of the states we have considered. All of the states are taken to have $\angles{\NumberOperator} = 8.0$. The states (a)-(h) are 2-spaced [(d) is approximately 4-spaced], while (i)-(l) are explicitly 4-spaced [(i) is $\infty$-spaced]. The states read: (a) squeezed vacuum, (b) even cat state, (c) number-phase state with $N=2$, (d) finite-energy GKP state, (e)-(h) moon states with $\FockScale = 1.0, 2.0, 3.0, 4.0$, consecutively from left to right, (i) Fock state $\ket{8}$, (j) Fock state superposition $(\ket{6}+\ket{10})/\sqrt{2}$, (k) compass state (or multi-legged/multi-headed/multicomponent cat state, depending on the convention), and (l) number-phase state with $N=2$. \label{FigS3}}
\end{figure*}

\begin{figure*}[ht!]
    \includegraphics[width=\linewidth]{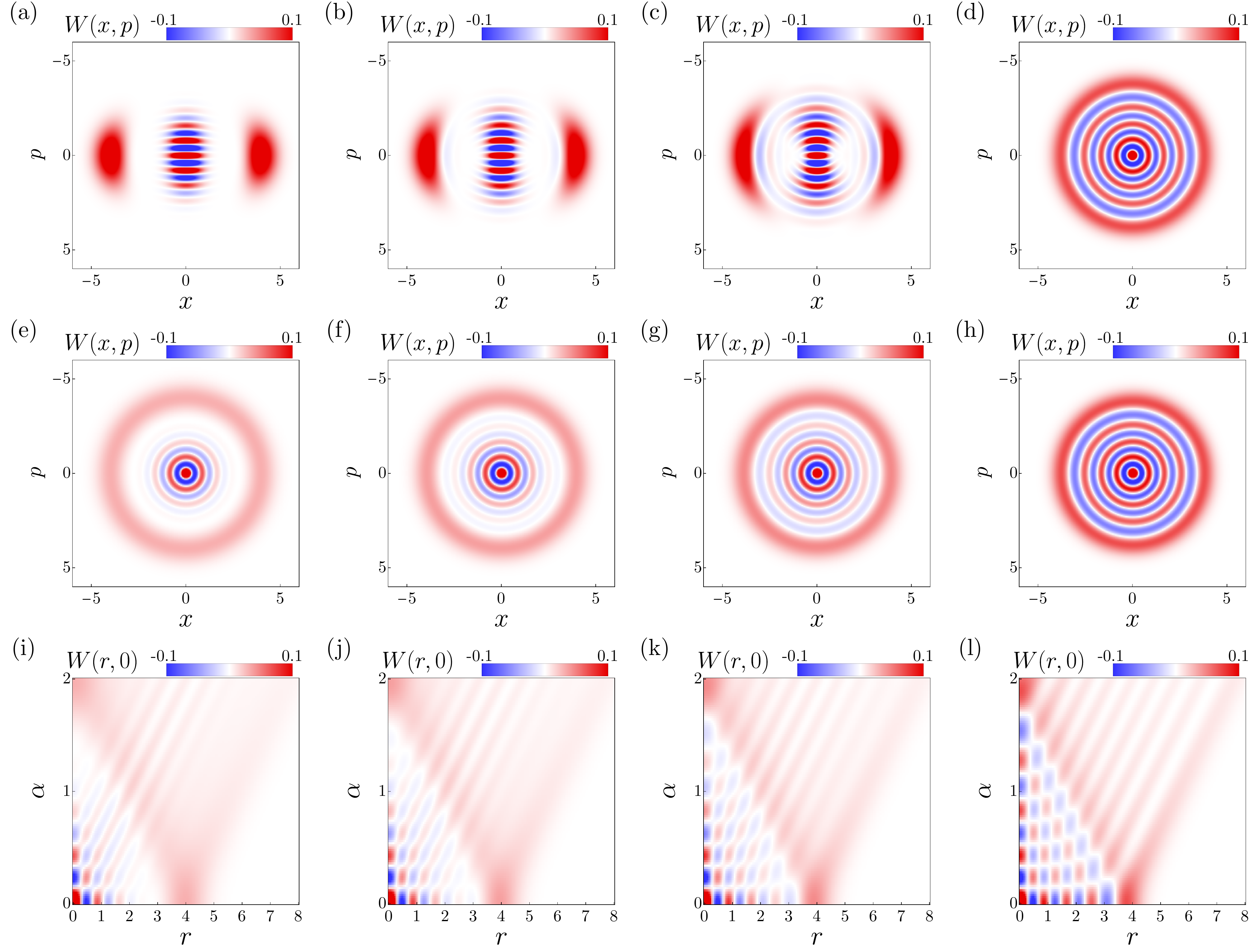}  
    \caption{Illustration of the effect of phase-randomized phase-space displacement. (a)-(d) Moon states with $\FockScale = 1.0, 2.0, 4.0, \infty$ from left to right. (d) represents perfect Fock state $\ket{8}$. (e)-(h) Corresponding states $\DensityMatrix_{\DispLetter=0}^{i}$ from Eq. (1) in the main text with additional phase noise such that only diagonal elements $\Prob_n^{\DispLetter}$ from Eq. (3) in the main text survive. The presented displacement is $\PhaseSpaceDispAmp = 0$, so only the effect of phase randomization is visible. Note the rotational symmetry of the states and that Wigner negativity survives the randomization. (i)-(j) Cuts of the Wigner functions of phase-randomized states with nonzero displacement $\PhaseSpaceDispAmp$ as a function of $\PhaseSpaceDispAmp$ and $r = \sqrt{x^2+p^2}$. Note that small phase-space features survive both phase randomization and some amount of displacement. \label{FigS3a}}
\end{figure*}

In Fig.~\ref{FigS3} we plot Wigner functions of all the states that we have considered, while in Fig.~\ref{FigS3a} we show the effect of phase-randomized phase-space displacement on the moon states.

\section{Quantum non-Gaussian states under decoherence in the small-time limit}
\begin{figure}[ht!]
    \includegraphics[width=\linewidth]{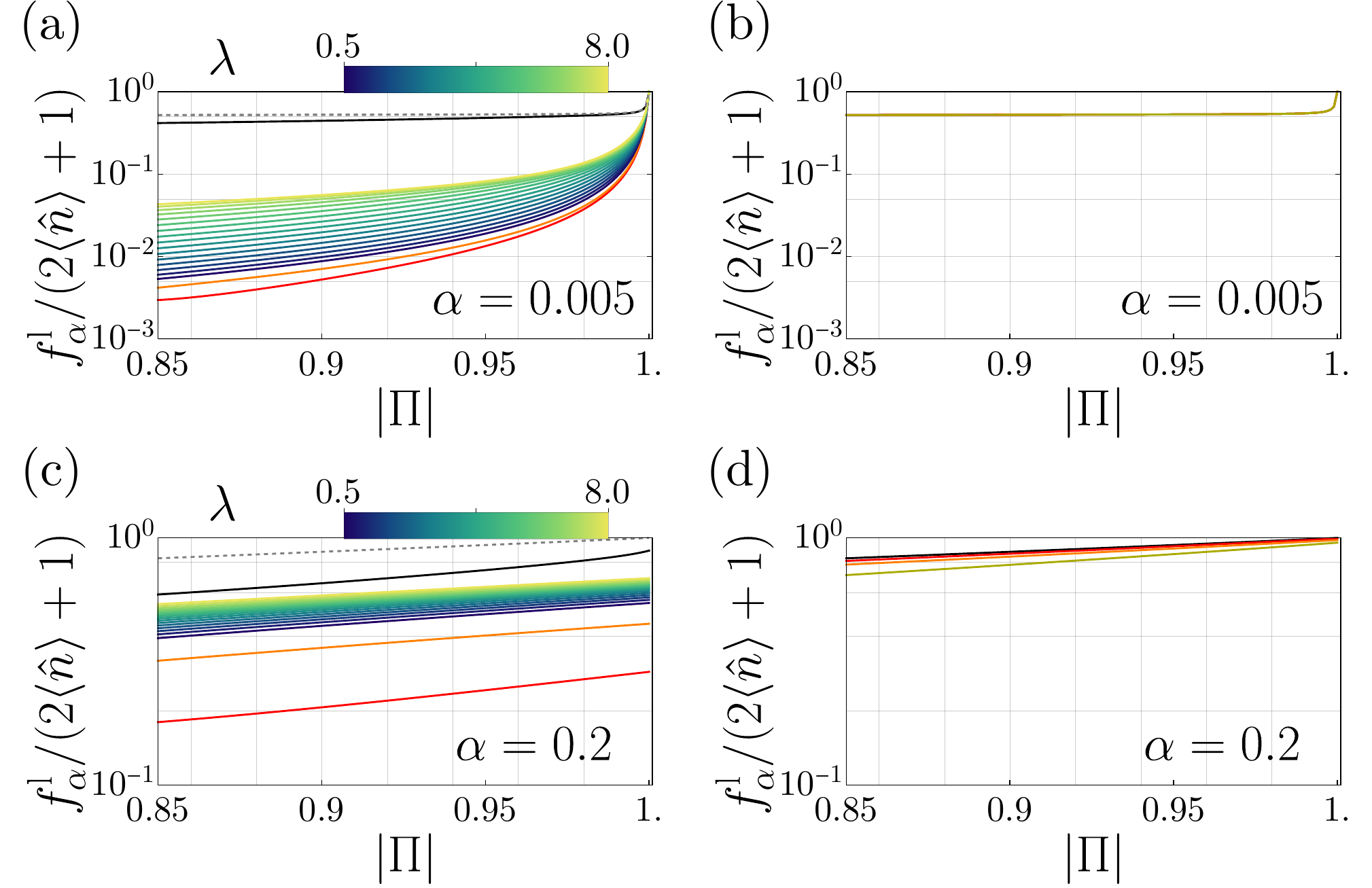}  
    \caption{Effect of the small-time loss on different quantum non-Gaussian states with $\angles{\NumberOperator} = 8.0$ for two different values of the force amplitude, $\PhaseSpaceDispAmp=0.005$ [(a) and (c)] and $\PhaseSpaceDispAmp=0.2$ [(b) and (d)]. We present the states shown in~Fig.~\ref{FigS3}: For (a) and (c) we have: the moon states (blue-green-yellow shades depending on $\FockScale$), Gaussian state (red solid line), 2-spaced number-phase state (orange solid line), finite-energy GKP state (black solid line), Fock state (gray dashed line); while for (b) and (d): Fock state (black solid line), $(\ket{6}+\ket{10})/\sqrt{2}$ (red solid line), compass state (orange solid line), 4-spaced number-phase state (yellow solid line). \label{FigS4}}
\end{figure}

\begin{figure}[h!]
    \includegraphics[width=\linewidth]{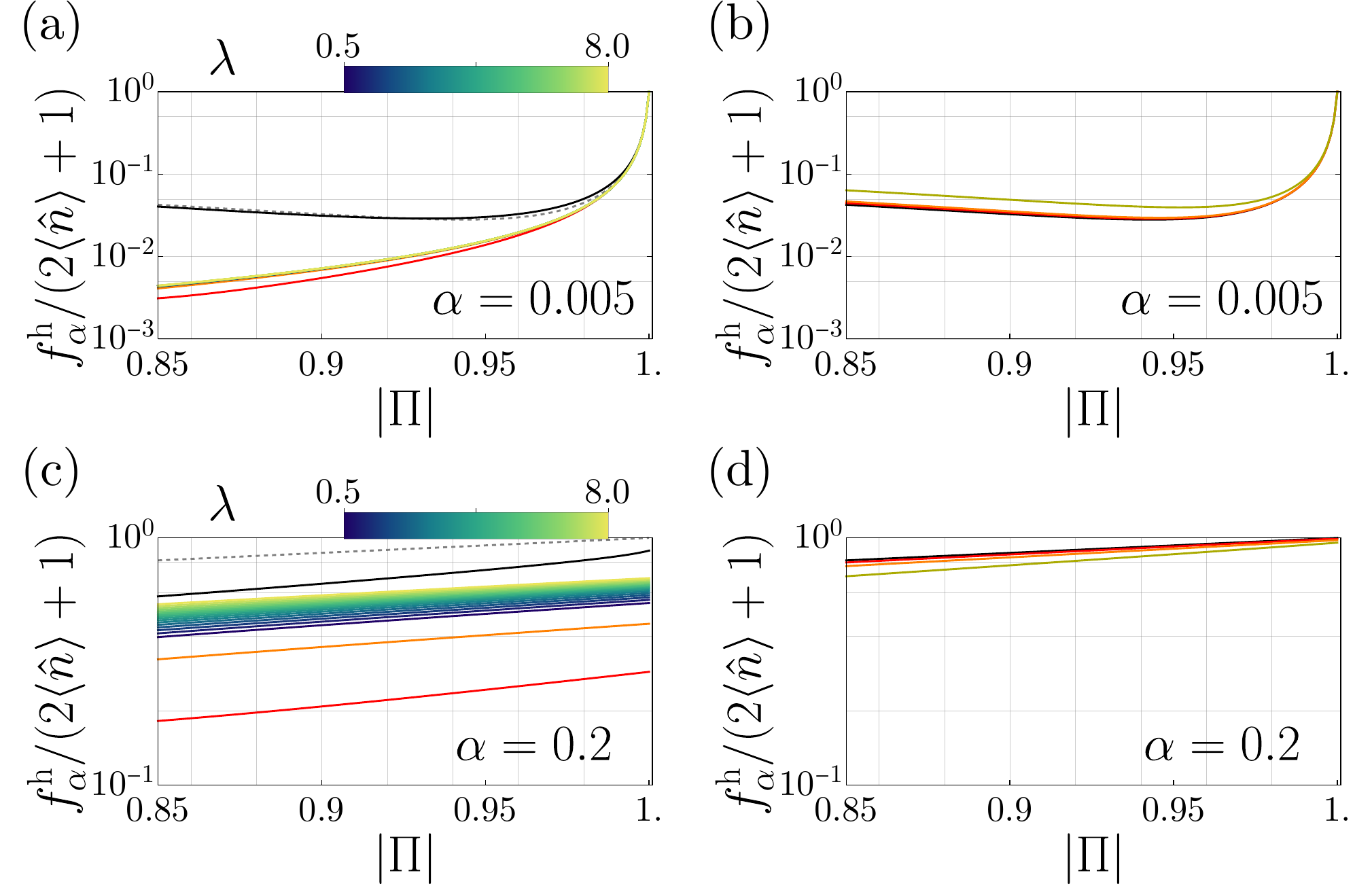}
    \caption{Effect of the small-time heating on different quantum non-Gaussian states. The states and labels are the same as in Fig.~\ref{FigS4}.  \label{FigS5}}
\end{figure}

Here we consider how the Fisher information of families defined in Sec.~\ref{supp:families} are affected by the small-time loss and heating given by~\eqref{ElementsDecoherence}.
To that end, we analyze states with $\angles{\NumberOperator} = 8.0$.
We use $\DevParity$ that quantifies deviation from ideal parity, $|\Parity| = |\text{Tr}[\rounds{-1}^{\NumberOperator} \DensityMatrix]| = 1 - \DevParity$, which evaluates to $\DevParity = 2 \angles{\NumberOperator} \SmallTime$ for loss and $\DevParity = 2 (2\angles{\NumberOperator}+1) \SmallTime \ThermalOccupation$ for heating.
The effect of loss is shown in~Fig.~\ref{FigS4} and the effect of the heating is shown in~Fig.~\ref{FigS5}.
In general, we find that within a decoherence model given by~\eqref{ElementsDecoherence}, for the loss, 4-spaced states are affected the same way as Fock states for small $\PhaseSpaceDispAmp$.
It is due to the facts that, first, within the model, the loss affects only $\Prob_{n-1}$ if $\Prob_{n}$ was nonzero (and 4-spaced states become 2-spaced ones, while 2-spaced ones become 1-spaced and lose metrological advantage), and second, the small value of $\PhaseSpaceDispAmp$ does not affect the value of the Fisher information.
It can be easily seen in~Fig.~\ref{FigS4}(b), where all the considered 4-spaced states exhibit the scaling of the Fock state.
For the 2-spaced states, the scalings are different [cf.~Fig.~\ref{FigS4}(a)] even for small $\PhaseSpaceDispAmp$.
There, we find that Gaussian states are affected the strongest, while finite GKP states the weakest (however, they are approximately 4-spaced).
The number-phase state fares better than the Gaussian one, and the moon states get progressively better as $\FockScale$ and they get closer to the Fock state.
For $\PhaseSpaceDispAmp = 0.2$ we observe the same behavior for 2-spaced states, while the 4-spaced states start to have distinguishable scaling---due to their finite spread in occupation number $\Delta^2 \NumberOperator$.
The Fock state is the most robust, followed by Fock superposition, compass, and number-phase, respectively.
\begin{figure}[ht!]
    \includegraphics[width=\linewidth]{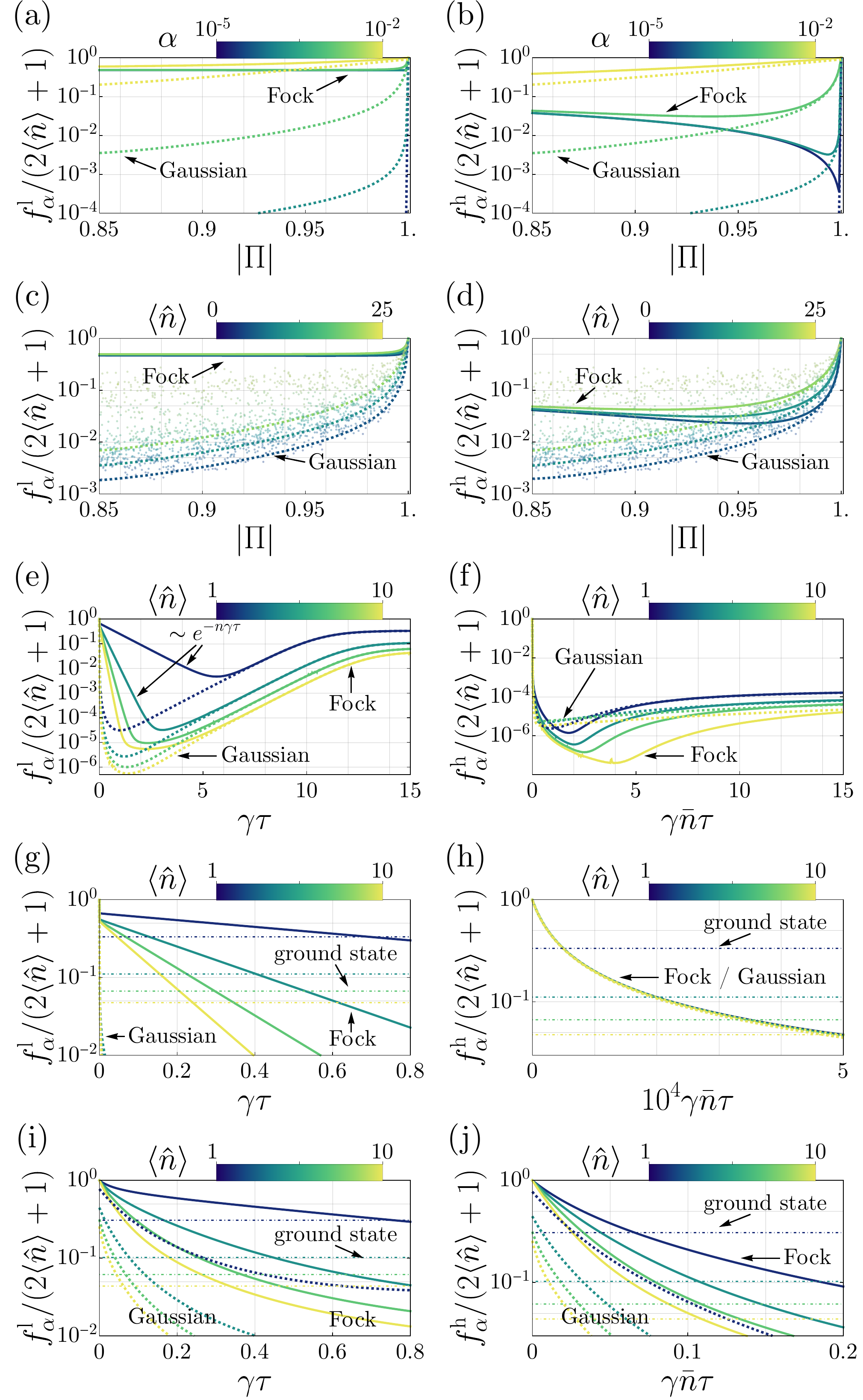}  
    \caption{Fisher information $\fisher_{\PhaseSpaceDispAmp}$ for Fock (solid lines) and Gaussian (dashed lines) states affected by excitation loss [label l, subplots(a), (c), (e), and (g)] or heating [h, (b), (d), (f), and (h)]. (a)-(b) $\fisher_{\PhaseSpaceDispAmp}$ for different values of $\PhaseSpaceDispAmp$ with fixed $\angles{\NumberOperator}=10$. In both cases, Fock states provide higher protection against the decoherence, despite the same value of $\fisher_{\PhaseSpaceDispAmp}$ in the decoherence-less scenario. (c)-(d) $\fisher_{\PhaseSpaceDispAmp}$ for different values of $\angles{\NumberOperator}$ with fixed $\PhaseSpaceDispAmp = 0.005$. Dots correspond to random states from Fig~\ref{Fig2}(a). Again, Fock states outperform the Gaussian states. (e)-(f) Full disspative dynamics driven by Master equation~\eqref{fullHamiltonian} for $\ThermalOccupation = 0$ (e) and large $\ThermalOccupation$ (f) with fixed $\PhaseSpaceDispAmp = 0.005$. The shown time scale is such that the final steady state is achieved in both cases. In the excitation loss case, Fock states always outperform the associated Gaussian states. For the heating case, short time dynamicse shows similar $\fisher_{\PhaseSpaceDispAmp}$ values for both Fock and Gaussian states, while for later times first Fock and then Gaussian states dominate. (g)-(h) Zoom into short time dynamics, a time scale over which sensing advantage over the ground state is lost. For excitation loss, Fock states largely outperform Gassian states, while for heating the difference is negligible. (i)-(j) The same as (g)-(h) but for $\PhaseSpaceDispAmp = 0.1$. For the heating case, the Fock states outperform Gaussian ones at the relevant time scale.  \label{FigS6}}
\end{figure}

The heating case presented in~Fig.~\ref{FigS5} is less favorable for 4-spaced states, as for a given nonzero $\Prob_{n}$, both $\Prob_{n-1}$ and $\Prob_{n+1}$ are affected.
Then, contrary to the effect of loss, 4-spaced states become 1-spaced ones, losing the metrological advantage.
First, we see that for $\PhaseSpaceDispAmp=0.2$, the decrease in the Fisher information has essentially the same behavior for both loss and heating [compare sublopts (c) and (d) of both~Fig.~\ref{FigS4} and~Fig.~\ref{FigS5}].
The behavior for $\PhaseSpaceDispAmp = 0.005$ is starkly different.
The moon states always have the same scaling, independent of $\FockScale$, and similar to the 2-spaced number-phase state (which resembles anti-squeezed moon state in our nomenclature).
The Gaussian state again has the worst scaling.
Peculiarly, there is a regime where a finite GKP state fares better than the Fock state.
The loss of the Fock state supremacy is further confirmed by looking at 4-spaced states.
Each of these states, including Fock superposition, compass, and 4-spaced number-phase state, has a better resilience against heating, with the last proving the most robust.

\section{Fock vs Gaussian}
In the main text we focused on the comparison between Fock and Gaussian states.
In Fig.~\ref{FigS6} we provide an extended version of Fig.~2 from the main text, covering analysis of different values of $\PhaseSpaceDispAmp$ and longer exact dynamics under decoherence.
The caption of the figure provides the analysis and comments.

\begin{figure}[ht!]
    \includegraphics[width=\linewidth]{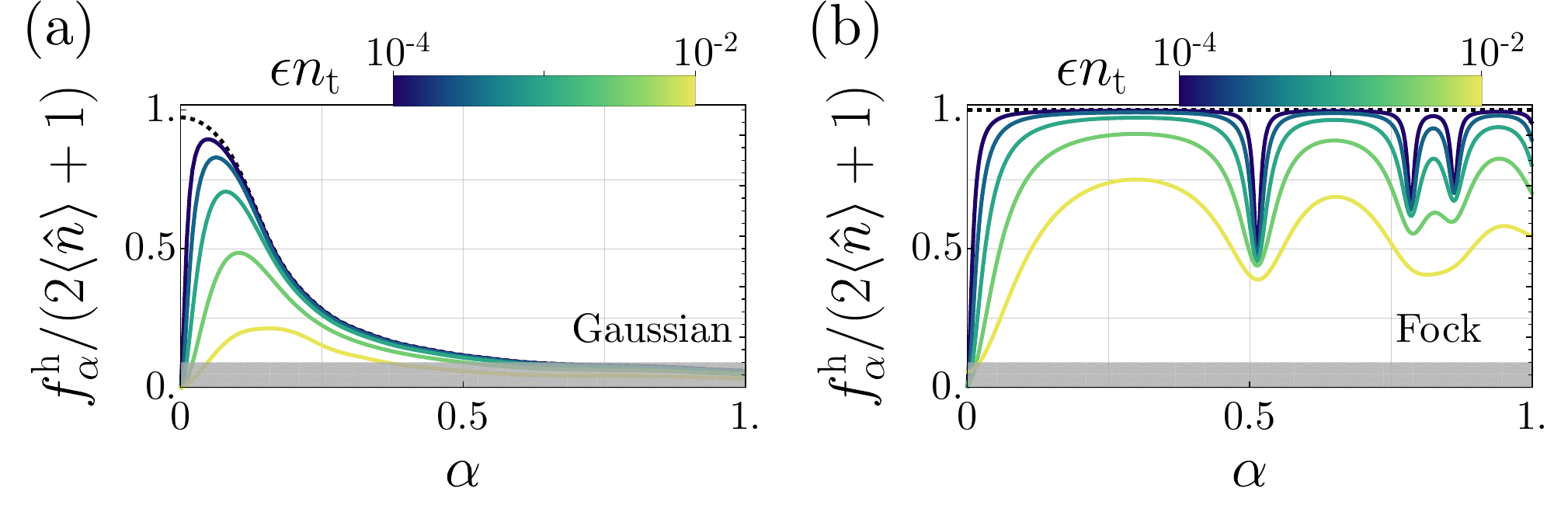}  
    \caption{ $\fisher_{\PhaseSpaceDispAmp}^{\HeatLetter}$ for Gaussian and Fock states with $\angles{\NumberOperator}=5$ at finite $\SmallTime \ThermalOccupation$, plotted against $\PhaseSpaceDispAmp$. The darker shaded region shows ground-state sensitivity; dashed line marks $\SmallTime \ThermalOccupation = 0$. Fock states exhibit metrological gain across the shown $\PhaseSpaceDispAmp$ range, unlike Gaussian ones. \label{FigS7}}
\end{figure}

\begin{figure}[ht!]
    \includegraphics[width=\linewidth]{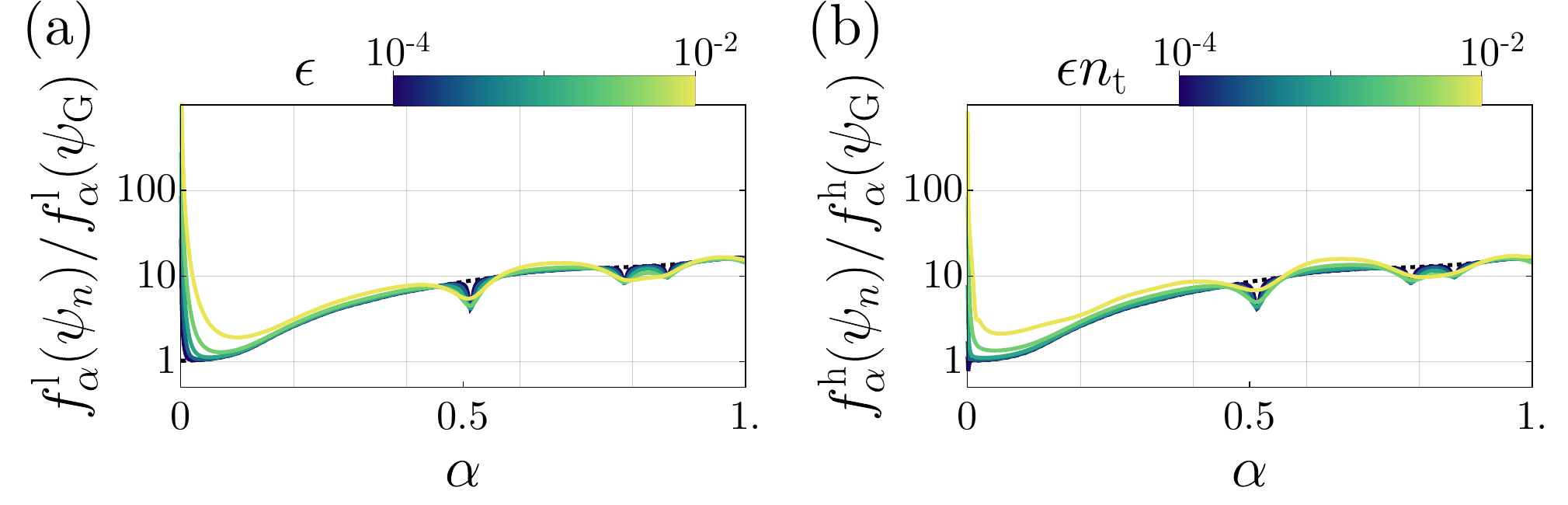}  
    \caption{ The ratio of the FI for Fock and Gaussian states, $\fisher_{\PhaseSpaceDispAmp} (\WaveFunction_n) / \fisher_{\PhaseSpaceDispAmp} (\WaveFunction_\text{G})$ both for loss and heating cases. The Fock states outperform the Gaussian states. The ratio first is much larger than 1 for very small $\PhaseSpaceDispAmp$, followed by the dip where the maximum of FI for the Gaussian state occurs, and then grows with increasing $\PhaseSpaceDispAmp$.  \label{FigS8}}
\end{figure}

Additionally, in Fig.~\ref{FigS7} we show the dynamical ranges in the heating case for Fock and Gaussian states compared to the ground-state sensitivity. In Fig.~\ref{FigS8} we show the ratio between the FIs for Fock and Gaussian states, showing how Fock states beat the limit set by Gaussian states.

\section{Additional details about the transition}
\begin{figure}[t!]
    \includegraphics[width=\linewidth]{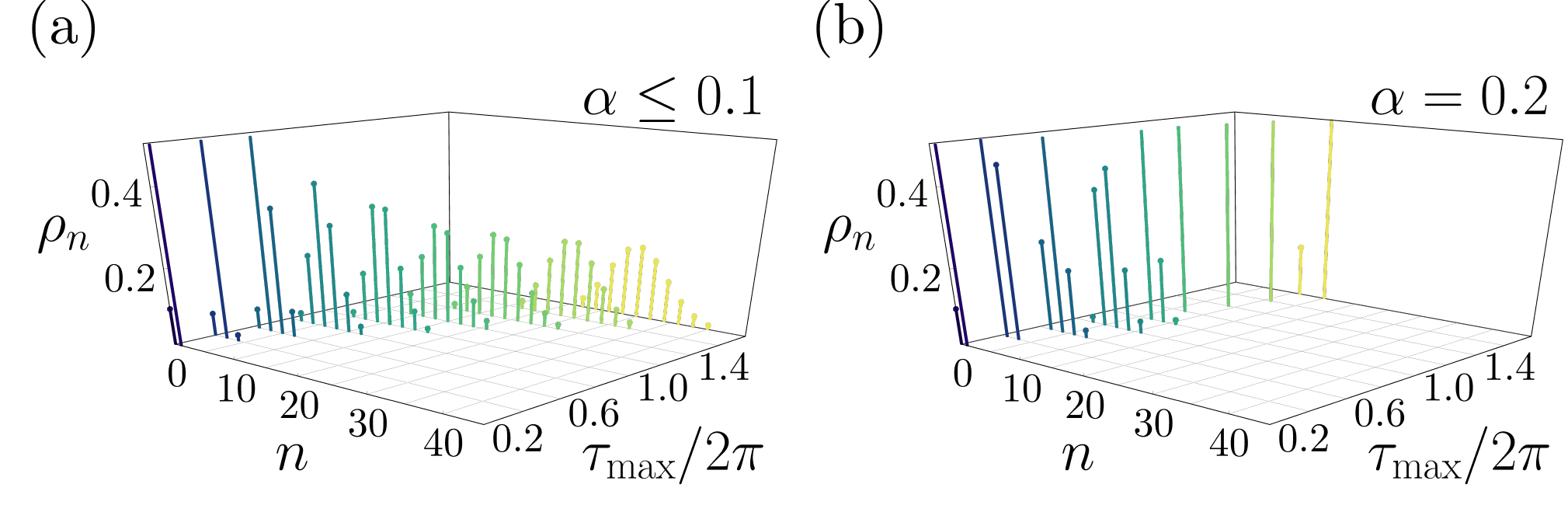}  
    \caption{Number distributions of the optimal solutions from Fig. 3 in the main text.
 \label{FigS10}}
\end{figure}

In Fig.~\ref{FigS10}, we additionally plot the number distributions of the probe states coming from the optimization procedure.

\section{Numerical details of dynamics and optimization}
For the solution of dynamics, both coherent and noisy, we use the QuTiP~\cite{Johansson2012,Johansson2013,Lambert2024} library. 
For each optimization run, we run dynamics with the least permissible size of the Fock space to speed up the computations.
After this part, we check that the results are indeed converged.
For control optimization, we use the Quantum Optimal Control Suite~\cite{Rossignolo2023} with the dCRAB~\cite{Muller2022} representation of the pulse and Nelder-Mead gradient-free direct search.
The control functions are expanded into
\begin{align}
    \PosControl(\Time) =\ScalingFunction(\Time) \sum_{k=1}^{N_{\text{p}}} a_k \cos \nu_k \Time + b_k \sin \nu_k \Time,
\end{align}
with the scaling function $\ScalingFunction(\Time)$,
\begin{align}
    f(\Time) = \tanh \squares{\ScalingParameter \sin\left(\pi \frac{\Time}{2 \TimeMax}\right)  } \tanh \squares{\ScalingParameter  \sin\left(\pi \frac{\TimeMax-\Time}{2 \TimeMax}\right)  },  
\end{align}
where $a_k$ and $b_k$ are parameters to be optimized, $\nu_k$ are frequencies that are probabilistically drawn from the uniform distribution in the range $[0,\nu_\text{max}]$, $\nu_\text{max}$ is the frequency cutoff, $N_{\text{p}}$ is the number of frequency components, and $\ScalingParameter = 10$ is a scaling parameter for the scaling function.
We take $N_{\text{p}} = 12$, $\nu_\text{max} = 50 \TimeMax / 2 \pi$, the maximal amplitude of the pulse $\text{max}_{\Time} |\PosControl(\Time)| = 150 $, and maximal number of evaluations in the super-iteration to be 10000.
The usual optimization runs lasted from minutes to hours and were run on a personal laptop.
\newpage

\end{document}